\pdfoutput=1 

\RequirePackage{snapshot} 
\ProvideExpandableDocumentCommand \class {} {sn-jnl}

\IfFileExists{sty/bootstrap.sty}{
  \RequirePackage{sty/bootstrap}
  \AppendInputPath{{sty/}{sty/img/}{fig/}{build/}}
}{
  \RequirePackage{bootstrap}
}

\ProvideIf[true]{arxiv} 
\ifarxiv
  \PassOptionsToClass {final} \class
  \PassOptionsToPackage {branded=false} {sn-jnl-setup}
\else
  \PassOptionsToClass {referee} \class
\fi

\RequirePackage{sn-jnl-setup}
\documentclass[  pdflatex,
  sn-mathphys,
  remarkboxoff,
  final,
] \class

\AddToHook{cmd/@maketitle/before}{\vspace{-3em}}
\def\Authorfont{\normalfont\fontsize{10bp}{12.5bp}\selectfont\boldmath\titraggedcenter}\def\addressfont{\normalfont\fontsize{10bp}{12.5bp}\selectfont\titraggedcenter}  

\IfClassLoadedWithOptionsTF \class {final} {} {\overfullrule=2cm}
\usepackage[bibtex={style=none}]{boilerplate}

\IfFileExists{main.reduced.bib}{
  \addbibresource{main.reduced.bib}
}{
  \IfFileExists{short.bib}{
    \addbibresource{short.bib}
  }{}
  \IfFileExists{bib.bib}{
    \addbibresource{bib.bib}
  }{}
  \IfFileExists{references2.bib}{
    \addbibresource{references2.bib}
  }{}
}

\usepackage{longtable}

\usepackage[capitalize]{cleveref}
\crefname{equation}{}{}
\Crefname{equation}{Equation}{Equations}

\usepackage{balance}

\usepackage[final]{simplediff}

\ProvideIf[false]{qtikz}

\usepackage{tikz}
\usetikzlibrary{arrows,arrows.meta}
\usepackage{tikzpagenodes}
\usepackage{qcircuit}
\usetikzlibrary{math}

\usetikzlibrary{math}
\usetikzlibrary{fpu}
\pgfkeys{/pgf/fpu}
\pgfkeys{/pgf/fpu/output format=fixed}
\tikzmath{
  function fold(\p, \q) { return \p*(1-\q) + \q*(1-\p); };
  function Hbin(\p) {
    if \p == 0 || \p == 1 then {
      return 0;
    } else {
      return - \p * log2(\p) - (1 - \p)*log2(1-\p);
    };
  };
  function Ibsc(\p, \q) { return Hbin(fold(\p, \q)) - Hbin(\q)); };
  function Ibec(\p, \q) { return Hbin(\p) - \q * Hbin(\p); };
  function IgaussC(\s) { return log2(1 + \s); };
  function IgaussR(\s) { return IgaussC(\s)/2; };
  function IzChannel(\p, \d) { return Hbin(\p * \d) - \p * Hbin(\d); };
  function zChannelPopt(\d) {
    real \g;
    let \g = (1-\d)^((1-\d)/\d);
    return \g/(1 + \d*\g);
  };
  function qHthermal(\n) {%
    if \n then { 
      return (\n + 1) * log2(\n + 1) - \n * log2(\n);
    } else { %
      return 0;
    };
  };
}
\pgfkeys{/pgf/fpu=false}
 
\IfFileExists{local-shorthands.tex}{

\NewDocumentCommand{\defname}{om}{\emph{#2}\IfValueT{#1}{~(#1)}}
\let\trace\undefined
\DeclareMathOperator{\trace}{Tr}

\DeclareGroup{\ket} |\rangle
\DeclareGroup{\bra} \langle|
\DeclareDocumentCommand{\braketTwo}{omm}{
  \IfValueTF {#1}
    { \sprod{ #2 #1| #3 } }
    { \sprod{ #2 \middle| #3 } }
}
\DeclareDocumentCommand{\ketbraTwo}{omm}{
  \IfValueTF {#1}
    { \ket[#1]{#2}\hspace{-0.6ex}\bra[#1]{#3} }
    { \ket{#2}\hspace{-0.6ex}\bra{#3} }
}
\DeclareDocumentCommand{\braket}{om}{
  \IfValueTF {#1}
    { \braketTwo[#1]{#2}{#2} }
    { \braketTwo{#2}{#2} }
}
\DeclareDocumentCommand{\ketbra}{om}{
  \IfValueTF {#1}
    { \ketbraTwo[#1]{#2}{#2} }
    { \ketbraTwo{#2}{#2} }
}

\newcommand{\identity}{\One}

\ExplSyntaxOn 
\clist_map_inline:nn {a,b,e,n} { \cs_new:cpn { h#1 } {\hat{#1}} }

\cs_new:Nn \set_more_upper_fonts:n {
  \cs_new:cpn { #1 set } {\use:c { c#1 }}
}
\int_step_inline:nn {\int_from_alph:n {Z}} { \set_more_upper_fonts:n {\int_to_Alph:n {#1}} }
\ExplSyntaxOff

\def\Rpool{R_{\text{pool}}}

\def\Xpool{F}

\def\Rbin{\tilde{R}}

\makeatletter
\DeclareDocumentCommand{\jorsn@enc@template}{O{Q}moo}%
  {\IfValueF{#3}{#2{#1}}{#2{#1}_{#3}}\IfValueT{#4}{\tup{#4}}}
\let\encTpl\jorsn@enc@template
\def\enc{\jorsn@enc@template\empty}
\def\encAlt{\jorsn@enc@template\tilde}

\def\Enc{\jorsn@enc@template\bm}
\def\EncAlt{\jorsn@enc@template[\tilde{Q}]\bm}

\makeatother

\def\err{e}
\def\erra{\err_{1}}
\def\errb{\err_{2}}

\let\saerr\aerr
\def\saerrya{\saerr_{1,1}}
\def\saerryb{\saerr_{1,2}}
\def\saerrza{\saerr_{2,1}}
\def\saerrzb{\saerr_{2,2}}

\def\codet{\mathcal{C}}

\def\CodeBL{\cB}

\let\type\hat

\def\capT{\mathsf{C}_{\mathsf{T}}}

\def\capID{\mathsf{C}_{\mathsf{ID}}}
\def\capSAID{\mathsf{C}_{\mathsf{ID}}}

\newcommand{\channel}{\mathcal{N}}

\newcommand{\ie}{\emph{i.e.} }
\newcommand{\eg}{\emph{e.g.} }
\newcommand{\etal}{\emph{et al.} }

}{}

\tikzset{
  font = {\figurecaptionfont\sffamily},   rateRegion/identification/.style = {color=darkgray},
  rateRegion/idTimeDivision/.style = {color=gray},
  rateRegion/transmission/.style   = {color=black!05},
  > = latex
}

\title{\baselineskip=0.7\baselineskip{}Identification Over Quantum~Broadcast~Channels}

\makeatletter
\hypersetup{
  ,unicode,pdfencoding=unicode,psdextra   ,colorlinks
  ,linkcolor={blue!80!black}
    ,citecolor={green!30!black}
  ,urlcolor={blue!80!black}
  ,pdfauthor={Uzi Pereg, Johannes Rosenberger, and Christian Deppe}
  ,pdftitle=\string\@title
  ,pdflang={English}
}
\makeatother

\begin{document}

\author*[1]{\fnm{Johannes}~\sur{Rosenberger}\orcidlink{0000-0003-2267-3794}}
  \email{johannes.rosenberger@tum.de}

\author[1]{\fnm{Christian}~\sur{Deppe}\orcidlink{0000-0002-2265-4887}}
  \email{christian.deppe@tum.de}

\author[1,2,3]{\fnm{Uzi}~\sur{Pereg}\orcidlink{0000-0002-3259-6094}}
  \email{uzipereg@technion.ac.il}

\affil*[1]{\orgname{Technical~University~of~Munich},
  \orgaddress{\country{Germany}};
  \orgdiv{School of Computation, Information and Technology, Department of Computer Engineering}}
\affil[2]{\orgname{Munich Center for Quantum Science and Technology},
  \orgaddress{\city{Munich},~\country{Germany}}}
\affil[3]{\orgname{Technion~–~Israel~Institute~of~Technology},
  \orgaddress{\city{Haifa}, \country{Israel}};
    \orgdiv{Helen Diller Quantum Center} and \orgdiv{Department of Electrical and Computer Engineering}}

\EarlyAcknow{
The authors wish to thank Roberto Ferrara (Technical University of Munich) for useful discussions.
This work was supported by the German Ministry of Education and Research (BMBF)
through grants
16KISQ028 (Uzi Pereg, Christian Deppe), 16KISK002 (Johannes Rosenberger, Christian Deppe), 16KIS1005 (Christian Deppe). Uzi Pereg was also supported by
German Research Foundation (DFG) under Germany’s Excellence Strategy – EXC-2111 – 390814868, and by 
the Israel CHE Fellowship for Quantum Science and Technology.
Christian Deppe was also supported by the Bavarian Ministry of Economic Affairs,
Regional Development and Energy in the project 6G and Quantum Technologies (6G QT).
}

\abstract{
Identification over quantum broadcast channels is considered. As opposed to the
information transmission task, the decoder only identifies whether a message of
his choosing was sent or not.
This relaxation allows for a double-exponential code size.
An achievable identification region is derived for a quantum broadcast channel, and 
a full characterization for the class of classical-quantum broadcast channels.
The identification capacity region of the single-mode pure-loss
bosonic broadcast channel
is obtained as a consequence.
Furthermore, the results are demonstrated for the quantum erasure broadcast channel,
where our region is suboptimal, but improves on the best previously known bounds.
}


\maketitle

\changelog

\section{Introduction}

Modern data systems have an ever-growing gap between the available information
storage and the bit-per-second rates, which are limited by the noisy
transmission medium~\cite{SilvaRodriguesAlbertiSolicAquino:17p}.
Quantum communication is thus expected to enter  the sixth generation of
cellular networks (6G) in order to achieve performance
gains~\cite{TariqKhandakerWongBennisDabbah:20p,FettwisBoche:21m,DangAminShihadaAlouini:20p}.

Data volumes are even larger when limiting a system to identifying alerts,
rather than recovering information. In Shannon's transmission task~\cite{shannon1948it0},
a transmitter sends a message over a noisy channel,
and the receiver needs to find which message was sent. 
In some modern event-triggered applications, however, the receiver may simply perform a
binary decision on whether a particular message of interest was sent or not.
This setting is known as identification  via channels~\cite{ahlswedeDueck1989id1}.
Identification (ID) is relevant for various applications such as
watermarking~\cite{steinbergMerhav2001id_watermarking,moulinKoetter2006id_watermark,ahlswedeCai2006watermarking}
and sensor communication~\cite{guenlueKliewerSchaeferSidorenko2021id_codes}.
In vehicle-to-X communication~\cite{bocheDeppe2018secureId_wiretap},
a vehicle may announce information about its future movements to the surrounding
road users. Every road user is interested in one specific movement
that interferes with its plans, and it checks only if this
movement is announced or not.

The ID capacity of a classical-quantum channel was determined by
Löber~\cite{Loeber1999PhD} and  Ahlswede and Winter~\cite{AhlswedeWinter2002quantumID}
(see also~\cite{Winter:13b}).
The ID capacity turns out to have the same value as the transmission capacity
for most classical-input single-user channels that we know of.
However, the units are different.
Specifically, the ID code size grows doubly exponentially in the block length,
provided that the encoder has access to a source of randomness.
Thereby, identification codes achieve an exponential advantage in throughput
compared to transmission codes. 
This is attained by letting the encoding and decoding sets overlap.
General results for ID are surveyed in~\cite{ahlswede2021identification_probabilistic_models}.
Löber~\cite{Loeber1999PhD} considered a simultaneous identification scenario,
in which the same measurement is performed in order to perform identification
for multiple receivers. This is also relevant to a network that consists of
chains~\cite{DiadamoBoche:19a}.
In \cite{BocheDeppeWinter2019quantum}, Boche~\etal considered identification
over the classical-quantum channel under channel uncertainty and secrecy
constraints.
\deleted{
Hayden and Winter \cite{HaydenWinter:12p}  further considered identification of
quantum states and showed that  the code size  is  exponential in this case, as
in transmission.
}
For quantum-quantum channels, even the single-user identification capacity is
unknown so far, except for special channels~\cite{Winter:13b}.
In general, it can exceed the transmission capacity of a quantum channel~\cite{Winter:13b},
and was recently shown to exceed the simultaneous identification capacity~\cite{atifPradhanWinter2023quantum_softCover_identification_arxiv}.
For example, the transmission capacity and the simultaneous identification
capacity of the noiseless qubit channel are both one~\cite{wilde2017quantum_it_2,atifPradhanWinter2023quantum_softCover_identification_arxiv},
but the identification capacity of the noiseless qubit channel is 2,
and equals the entanglement-assisted transmission capacity~\cite{Winter:13b}.
The best lower bounds equal the amount of common randomness that can be generated
over a channel, and thus entanglement also increases the identification capacity~\cite{Winter:13b}.
 
The broadcast channel is a fundamental multi-user communication model, whereby a
single transmitter sends messages to two receivers~\cite{coverThomas2005elements}.
In the traditional transmission setting, the capacity region of the discrete
memoryless broadcast channel is generally unknown, even in the classical case.
The best known lower bound is due to Marton~\cite{marton1979DMBC},
and the best known upper bound was proven by
Nair and El Gamal~\cite{nairElgamal2007BC_outer_bound}.
The two bounds coincide in special cases such as more capable,
less noisy or degraded broadcast channel~\cite{elgamal1979BC_class}.
On the other hand, the ID capacity region of the classical broadcast channel was fully characterized by
Bracher and Lapidoth~\cite{bracherLapidoth2017idbc,bracher2016PhD},
for uniformly distributed messages.
Namely, the ID capacity region is known for \emph{any} classical
discrete memoryless broadcast channel, without special requirements on the channel.
The derivation in~\cite{bracherLapidoth2017idbc,bracher2016PhD}
is based on a pool-selection technique that differs from the standard arguments.
Related settings were also considered in~\cite{verbovenMeulen1990idbc,bilikSteinberg2001id_dbc,ahlswede2008gtid_updated}.
The authors of the present paper have recently considered ID over the classical
compound multiple-input multiple-output (MIMO) broadcast
channel~\cite{rosenbergerPeregDeppe2022id_compound_BC_conference,rosenbergerPeregDeppe2023id_compound_BC}.
 
Quantum broadcast channels were studied in various
settings~\cite{YardHaydenDevetak:11p,SavovWilde:15p,RadhakrishnanSenWarsi:16p,WangDasWilde:17p,
DupuisHaydenLi:10p,Dupuis:10z,HircheMorgan:15c,SeshadreesanTakeokaWilde:16p,BaumlAzuma:17p,
HeinosaariMiyadera:17p,BochCaiDeppe:15p,Hirche:12z,XieWangDuan:18c,Palma:19p,
AnshuJainWarsi:19p1,ChengDattaRouze:19a}.
Yard~\etal~\cite{YardHaydenDevetak:11p} derived the superposition inner bound
and determined the capacity region for the degraded classical-quantum broadcast
channel.
Wang~\etal~\cite{WangDasWilde:17p} used the previous characterization to
determine the capacity region for Hadamard broadcast channels.
Dupuis~\etal~\cite{DupuisHaydenLi:10p,Dupuis:10z} developed the
entanglement-assisted version of Marton's region.
Quantum broadcast channels with conferencing decoders were recently considered
in~\cite{PeregDeppeBoche:21p2} as well, providing an information-theoretic
perspective to the operation of quantum repeaters.
In addition, security aspects were treated in~\cite{SalekHsiehFonollosa:19a,SalekHsiehFonollosa:19c}.

Optical communication forms the backbone of the
Internet~\cite{BardhanShapiro:16p,Pereg:21c2,Pereg:21p,KumarDeen:14b}.
The Gaussian bosonic channel is a simple quantum-mechanical model for optical
communication over free space or optical fibers~\cite{WPGCRSL:12p,WildeHaydenGuha:12p}.
An optical communication system consists of a modulated source of photons, the
optical channel, and an optical detector. For a single-mode bosonic broadcast
channel, the channel input is an electromagnetic field mode with annihilation
operator $\ha$, and the output is a pair of modes with  annihilation operators
$\hb_1$ and $\hb_2$, corresponding to each receiver. Bosonic broadcast channels
are considered in different settings
in~\cite{GuhaShapiro:07c,GuhaShapiroErkmen:07p,DePalmaMariGiovannetti:14p,
TakeokaSeshadreesanWilde:16c,TakeokaSeshadreesanWilde:17p,LaurenzaPirandola:17p,
AndersonGuhaBash:21c,peregFerraraBloch2021key_secrecy_bosonicBC_itw}.

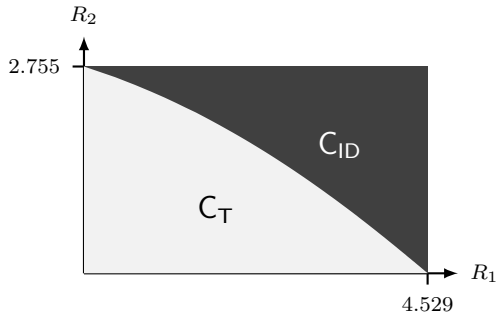
\begin{figure}[tb]
  \centering
  \providecolor{identification}{named}{blue}
\providecolor{idTimeDivision}{named}{orange}
\providecolor{transmission}{named}{red}

\newcommand{\rateRegionPureLossBosonicBC}[3][20]{

\pgfkeys{/pgf/fpu/output format=fixed}
\tikzmath{
 real \e, \NA, \Ra, \Rb;
 \e = #2;
 \NA = #3;
 function R2(\x) { return qHthermal((1 - \e) * \NA) - qHthermal((1 - \e) * \x * \NA); };
 function R1(\x) { return qHthermal(\e * \x * \NA); };
 \Ra = R1(1);
 \Rb = R2(0);
}
\pgfkeys{/pgf/fpu=false}

\begin{tikzpicture}[auto,inner sep=1ex, thick, node distance=2cm and 2cm, box/.style={draw,inner sep=1ex},scale=#1]
\pgfkeys{/pgf/number format=fixed}
\pgfkeys{/pgf/number format/precision=3}
\draw[->] (0,0) -- (0, {\Rb+0.4}) node[above] {$R_2$};
\draw[->] (0,0) -- ({\Ra+0.4}, 0) node[right] {$R_1$};

\fill[rateRegion/identification] (0,0) -- (0, \Rb) -- +(\Ra, 0) node[below left,color=white,xshift=-2em,yshift=-2em] {\large$\capSAID$} -- (\Ra, 0);
\draw (0, \Rb) -- +(-0.15,0) node [left] {$\pgfmathprintnumber{\Rb}$}; 
\draw (\Ra, 0) -- +(0,-0.15) node [below] {$\pgfmathprintnumber{\Ra}$}; 

\pgfkeys{/pgf/fpu/output format=fixed}
\fill[rateRegion/transmission] (0,0) -- plot [parametric,domain=0:1,range=0:1,samples=100] ({R1(\x)}, {R2(\x)});
\pgfkeys{/pgf/fpu=false}
\draw ({R1(0.2)}, {R2(0.2)}) node[below left,xshift=-1em,yshift=-1em] {\large$\capT$};
\end{tikzpicture}
}
  \def\etaTikz{0.8}
  \def\NATikz{10}
  \rateRegionPureLossBosonicBC[1]{\etaTikz}{\NATikz}
  \caption[ID and transmission capacity regions of the bosonic broadcast channel]{     \label{fig:bosonicBC}The transmission and ID capacity regions of the
      pure-loss bosonic broadcast channel, with coherent-state encoding,
      mean photon-number input constraint $N_A = \NATikz$,
     and transmissivity $\eta = \etaTikz$. The transmission capacity region $\capT$
     corresponds to the light gray area, and the ID capacity region $\capSAID$
     comprises additionally the dark gray rectangular area.}
\end{figure}

In this work, we consider identification over the quantum broadcast channel.
We derive an achievable ID rate region for the general quantum
broadcast channel, and establish full characterization for the 
classical-quantum broadcast channel under a semi-average error criterion. 
We demonstrate our results and determine the ID capacity region
of the quantum erasure broadcast channel.
Furthermore, we establish the capacity region of the single-mode
pure-loss bosonic broadcast channel with coherent-state encoding.
The ID capacity region of the bosonic broadcast channel is depicted in
Figure~\ref{fig:bosonicBC} as the area below the solid blue line.
For comparison, the transmission capacity region, as  determined
by Guha and Shapiro \cite{GuhaShapiro:07c} subject to the minimum
output-entropy conjecture, is indicated by  the red dashed line.
It can be seen that the ID capacity region is significantly larger
than the transmission counterpart. We note that the ID result does
not require the conjecture.

While the properties above are analogous to the classical setting \cite{rosenbergerPeregDeppe2023id_compound_BC}, the analysis is more involved.
To prove the direct part, we extend the pool-selection method due to Bracher and Lapidoth \cite{bracherLapidoth2017idbc,bracher2016PhD}
to the quantum setting.
On the other hand, our converse proof is based on completely different arguments than in Bracher and Lapidoth's classical proof. Instead, we exploit recent observations made by Boche et al. \cite{BocheDeppeWinter2019quantum} as they treated the classical-quantum compound channel, combined with the arguments of Ahlswede and Winter \cite{AhlswedeWinter2002quantumID} in their seminal paper on ID for the single-user classical-quantum channel.

This paper is organized as follows:
In Section~\ref{sec:prelim}, we introduce the notation, give basic definitions, and introduce
the communication model.
Section~\ref{sec:results} contains our main results.
In Section~\ref{sec:examples}, we demonstrate our results
for the pure-loss bosonic broadcast channel
and the erasure broadcast channel.
Section~\ref{sec:proof.capSAID-CBC.achiev} provides the
achievability proof for identification over the quantum broadcast channel in finite dimensions,
and Section~\ref{sec:proof.capSAID-CBC.converse} provides
the proof for the ID capacity region of the classical-quantum broadcast channel.
Finally, the results are summarized in Section~\ref{sec:summary}.

\section{Preliminaries and Related Work}
\label{sec:prelim}

\subsection{Notation}
\label{subsec:notation}

We use the following notation conventions.

\subsubsection{Basic Notation}

\deleted{
Script letters $\cX,\cY,\cZ,...$ are used for finite sets.
Lowercase letters $x,y,z,\ldots$  represent constants and values of classical random variables, and uppercase letters $X,Y,Z,\ldots$ represent classical random variables.  
The distribution of a  random variable $X$ is specified by a probability mass function (PMF) 
$P_X(x)$ over a finite set $\cX$. The set of all PMFs over $\cX$ is denoted by $\cP(\cX)$.  	
We use $x^j=(x_1,x_{2},\ldots,x_j)$ to denote  a sequence of letters from $\cX$. 
A random sequence $X^n$ and its distribution $P_{X^n}(x^n)$ are defined accordingly. 
}
\added
{
\begin{center}
\begin{longtable}{ll}
$X,Y,\dots$ & classical random variables \\
$\cX,\cY, \dots$ & finite sets (alphabets)\\
$x,y,\dots$ & constants and classical values \\  
        $x^n = (x_1,x_2,\dots,x_n) \in \cX^n$ & sequence of length $n$\\
  $P_X$ & probability mass function (PMF) of $X$\\
  $\expect[X]   $ 
  & expectation of a random variable $X$ \\
        $\cP(\cX)$ & set of all PMFs with finite support over a set $\cX$ \\
  $P^n(x^n) = \prod_{t=1}^n P(x_t)$ & $n$-fold product distribution \\
  $[N]$ & $\set{1, \dots, \ceil{N}}$ \\
                $A,B,\dots$ & quantum systems\\
  $\cH_A$  & Hilbert space  $A$ \\
  $\rho_A \in \scrD(\cH_A)$ & density operator on $\cH_A$ \\
  $\scrD(\cH_A)$ & set of density operators on $\cH_A$ \\
  $\mathcal{N}_{A\to B}:\scrD(\cH_A)\to \scrD(\cH_B) $& quantum channel (CPTP map)\\
    $\set{D_j: j\in [J]}$ & positive operator-valued measure (POVM), 
  \\
                          $\ket{\Phi_{AB}} = \frac 1 {\sqrt{d}} \sum\limits_{i=0}^{d-1} \ket i _A \otimes \ket i _B$
                    & a maximally entangled state of of dimension $d$
\end{longtable}
\end{center}
}

\added{\protect{
\subsubsection{Information Measures}
\begin{center}
\begin{longtable}{ll}
  $H(X) = \sum\limits_{x \in \supp P_X} -P_{X}(x) \log_2 P_X(x)$
    & classical entropy     \\
      $I(X; Y) =    H(X) + H(Y) - H(X Y)$
    & classical mutual information \\   $H(A)_\rho = H(\rho_A) = - \trace[\rho_A\log_2(\rho_A)]$
    & quantum entropy \\   $I(A; B)_\sigma = H(\sigma_A) + H(\sigma_B) - H(\sigma_{AB})$
    & quantum mutual information \\   $H(A|B)_\sigma = H(\sigma_{AB}) - H(\sigma_B)$
    & conditional quantum entropy
\end{longtable}
\end{center}
}}

\subsubsection{Quantum Broadcast Channels}
\label{subsec:Qchannel}

A quantum broadcast channel 
$\channel_{ A\rightarrow B_1 B_2}: \mathscr{D}(\cH_A)\to \mathscr{D}(\cH_{B_1}\otimes \cH_{B_2})$
corresponds to a quantum physical evolution from the input $A$ to the combined output $B_1,B_2$, associated with the transmitter and two
receivers, respectively.
We assume that the channel is memoryless. That is, if the systems
$A^n=(A_1,\ldots,A_n)$ are sent through $n$ channel uses, then the input
$\rho_{ A^n}$ undergoes the tensor product mapping
$\channel_{ A^n\rightarrow B_1^n B_2^n}\equiv  \channel_{ A\rightarrow B_1 B_2}^{\otimes n}$.
The marginal channel  is defined by
$
\channel_{A\rightarrow B_1}^{(1)}(\rho_A)=\trace_{B_2} \left( \channel_{ A\rightarrow B_1 B_2}(\rho_{A}) \right) 
$
for Receiver 1, and similarly $\channel_{A\rightarrow B_2}^{(2)}$ for Receiver 2. 
The transmitter, Receiver 1, and Receiver 2 are often called Alice, Bob 1, and Bob 2. 
A classical-quantum (c-q) broadcast channel $\channel^{\,\text{c-q}}_{X\to B_1 B_2}$ is defined, in a similar manner, as
a mapping $\cX \to \mathscr{D}(\cH_B)$.

\subsection{Identification Codes}
\label{sec:prelim:ID}

In the following, we define the communication task of identification over a quantum broadcast channel,
where the decoder is not required to recover the sender's message
\(i\), but simply determines whether a particular message \(i'\) was sent or not.
\begin{definition}
An \(\tup{N_1, N_2, n}\) identification (ID) code  for the
quantum broadcast channel \(\channel_{A\to B_1 B_2}\) consists of an encoding channel
\(\cE_{A^n}:[N_1]\times [N_2]\to \mathscr{D}(\cH_A^{\otimes n})\)
and a collection of binary decoding POVMs
\(\cD_{B_1^n}^{i_1}=\{\identity-D_{i_1}^{(1)},D_{i_1}^{(1)}\}\)
and \(\cD_{B_2^n}^{i_2}=\{\identity-D_{i_2}^{(2)},D_{i_2}^{(2)}\}\),
for \( i_1 \in [N_1]\) and \( i_2 \in [N_2]\).
We denote the identification code by
$\codet = (\cE_{A^n}, \cD_{B_1^n}, \cD_{B_2^n})$.

The identification scheme is depicted in Figure~\ref{fig:qbID}.
Alice chooses a pair of messages $(i_1,i_2)$, where $i_k\in [N_k]$, for $k\in\{1,2\}$.
She encodes the messages by preparing an input state $\rho^{i_1,i_2}_{A^n}\equiv \cE_{A^n}(i_1,i_2)$,
and sends the input system $A^n$ through $n$ uses of the quantum broadcast channel $\channel_{A\to B_1 B_2}$.
Bob 1 and Bob 2 receive the output systems $B_1^n$ and $B_2^n$, respectively.
Suppose that Bob \(k\) is interested in a particular message
\(i'_k \in [N_k]\), where $k\in\{1,2\}$.
Then, he performs the binary measurement $\{\identity-D_{i_k}^{(k)},D_{i_k}^{(k)}\}$ to determine whether
\(i_k'\) was sent or not, and obtains a measurement outcome $s_k\in\{0,1\}$.
He declares `no' if the measurement outcome is $s_k=0$, and `yes' if  $s_k=1$. 

\begin{figure*}
\newcommand{\bosonicBC}[9]{
\def\x{#1}
\def\y{#2_1}
\def\z{#2_2}

\def\channel{#3}

\def\decY{#4}
\def\decZ{#5}

\def\messageY{#6}
\def\messageZ{#7}
\def\resultY{#8}
\def\resultZ{#9}

\def\spacing{1}

\usetikzlibrary{positioning}

\noindent
\begin{tikzpicture}[auto,inner sep=1ex, thick, node distance=1cm and 1cm, box/.style={draw,inner sep=1ex}]

\node (source) {};
\node[right=1.3 of source,box] (Enc) {$\cE_{A^n}$};
\node[right=\spacing of Enc,box,minimum height=5em] (W) {$\channel$};

\node[left=1pt of W.north east,below,minimum height=1.75em,inner sep=0,outer sep=0] (WY) {};

\node[right=\spacing of WY,box] (DecY) {$\decY$};
\node[right=2.2 of DecY] (sinkY) {};

\node[left=1pt of W.south east,above,minimum height=1.75em,inner sep=0,outer sep=0] (WZ) {};
\node[right=\spacing of WZ,box] (DecZ) {$\decZ$};
\node[right=2.2 of DecZ] (sinkZ) {};

\path[draw]
    (source) edge[->] node[above] {$(\messageY, \messageZ)$} (Enc.west)
    (Enc.east) edge[->] node[above] (x) {$\x$} (W.west)

    (WY) edge[->] node[above] (y) {$\y$} (DecY.west)
    (DecY.east) edge[->] node[above] (mY) {\resultY} (sinkY)

    (WZ) edge[->] node[above] (z) {$\z$} (DecZ.west)
    (DecZ.east) edge[->] node[above] (mZ) {\resultZ} (sinkZ)

    (DecY.north) edge[<-] node[above,shift={(0,0.6ex)}] {$i_1'$} +(0,2.2ex)
    (DecZ.south) edge[<-] node[below,shift={(0,-0.6ex)}] {$i_2'$} +(0,-2.2ex)
;
\end{tikzpicture}
}
\centering
\sffamily
\bosonicBC{A^n}{B^n}
     {\cN^{\otimes n}}
     {\cD_{B^n_1}}
     {\cD_{B^n_2}}
     {i_1}{i_2}
     {Was $i_1'$ sent?}
     {Was $i_2'$ sent?}

\caption{Identification over the quantum broadcast channel $\channel_{A\to B_1 B_2}$. Alice chooses a  message pair $(i_1,i_2)$. She encodes the messages by preparing an input state $ \cE_{A^n}(i_1,i_2)$, and sends the input system $A^n$ through $n$ uses of the quantum broadcast channel $\channel_{A\to B_1 B_2}$.
Bob 1 and Bob 2 receive the output systems $B_1^n$ and $B_2^n$, respectively.
As Bob \(k\) is interested in the  message
\(i'_k \in [N_k]\),  he performs the binary measurement
$\cD_{B_k^n}^{i_k'} = \{\identity-D_{i_k'}^{(k)},D_{i_k'}^{(k)}\}$
to determine whether \(i_k'\) was sent or not. }
\label{fig:qbID}
        \end{figure*}
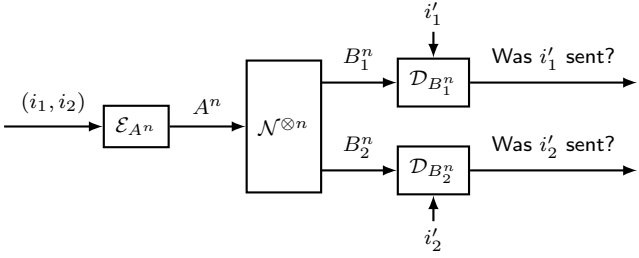

The ID rates of the code \(\codet\) are defined as
\(R_k = \frac{1}{n} \log\log(N_k)\), for \(k \in \set{1,2}\).
In this work, we assume that the ID messages \(i_k\) are uniformly distributed over
the set \([N_k]\), for \(k \in \set{1,2}\). Therefore, the error probabilities are defined on
average over the messages for the other receiver.
Bob 1 makes an error in two cases: (1) He decides that
\(i_1\) was \emph{not} sent (missed ID); (2) Bob 1 decides that \(i'_1\) was sent,
while in fact \(i_1\) was sent, and \(i_1 \neq i'_1\) (false ID). The probabilities of these two kinds of error,
averaged over \(i_2 \in [N_2]\),  are defined as
\begin{subequations}
\label{eq:def.SAID-errProbs}
\begin{align}
  \saerr_{1,1} (\channel, n, \codet, i_1)
    &= \frac{1}{N_2} \sum_{i_2\in [N_2]}
     \trace\intv{
       (\identity-D_{i_1}^{(1)})
       \channel^{\otimes n}
       (\cE(i_1,i_2))
     },
   \\
  \saerr_{1,2} (\channel, n, \codet, i'_1, i_1)
    &= \frac{1}{N_2} \sum_{i_2\in [N_2]}
     \trace\intv{
       D_{i'_1}^{(1)}
       \channel^{\otimes n}
       (\cE(i_1,i_2))
     },
\end{align}
for $i_1,i_1'\in [N_1]$ such that $i_1\neq i_1'$.
Similarly, Bob 2's error probabilities are
\begin{align}
  \saerr_{2,1} (\channel, n, \codet, i_2)
    &= \frac{1}{N_1} \sum_{i_1\in [N_1]}
     \trace\intv{
       (\identity- D_{i'_2}^{(2)} )\channel^{\otimes n}
       (\cE(i_1,i_2))
     },
     \\
  \saerr_{2,2} (\channel, n, \codet, i'_2, i_2)
    &= \frac{1}{N_1} \sum_{i_1\in [N_1]}
     \trace\intv{
       D_{i'_2}^{(2)}
       \channel^{\otimes n}
       (\cE(i_1,i_2))
     }.
\end{align}
for $i_2,i_2'\in [N_2]$ such that $i_2\neq i_2'$.
\end{subequations}

An \((N_1, N_2, n, \lambda_1,\lambda_2)\) ID-code \(\codet\)
for the quantum broadcast channel \(\channel_{A\to B_1 B_2}\)
satisfies
\begin{subequations}
\begin{align}
   \max_{i_k \in [N_k]} \saerr_{k,1}(\channel, n, \codet, i_k) &< \lambda_1, \\
 \max_{\substack{i_k,i'_k \in [N_k], \\ i'_k \neq i_k}}
     \saerr_{k,2}(\channel, n, \codet, i'_k, i_k) &< \lambda_2,
\end{align}
\end{subequations}
for \(k \in \set{1,2}\).
An ID rate pair \((R_1, R_2)\) is \emph{achievable} if for every \(\lambda_1,\lambda_2 > 0\) and
sufficiently large \(n\), there exists an
\(\tup{\exp{e^{nR_1}}, \exp{e^{nR_2}}, n, \lambda_1,\lambda_2}\) ID-code.
The ID capacity region \(\capSAID(\channel)\) of the quantum broadcast channel \(\channel_{A\to B_1 B_2}\)
is defined as the set of achievable rate pairs.
\end{definition}

\subsection{Previous Results}
\label{sec:relatedWork}
In the traditional transmission setting \cite{shannon1948it0}, the decoder Bob
is required to find an estimate $\hat{i}$ of Alice's message. This is a more
stringent requirement than identification and it results in exponentially slower
communication. Specifically, the number of messages  scales as $\exp(nR)$ for
transmission, whereas $\exp(e^{nR})$ for identification. While the transmission
rate is measured in units of information bits per channel use, the
identification rate has different units. Nonetheless, for the
classical-quantum single-user channel,
it turns out that the identification and transmission capacities have
the same \emph{value}.

In the single-user setting, the ID capacity of the classical-quantum channel was
determined by Löber~\cite{Loeber1999PhD} and Ahlswede~and~Winter~\cite{AhlswedeWinter2002quantumID}.
Let $\cW_{X\to B}$ be a single-user c-q channel. The ID capacity $C_{\text{ID}}(\cW)$ is then defined,
in a similar manner, as the supremum of achievable ID rates over the c-q channel $\cW_{X\to B}$. 

\begin{theorem}[see {\cite{Loeber1999PhD,AhlswedeWinter2002quantumID}{\cite[Theorem 4]{winter2004quantum}}}]
The ID capacity of a single-user classical-quantum channel \(\cW_{X\to B}\) is given by
\begin{gather}
  C_{\text{ID}}(\cW) = \max_{P_X \in \cP(\cX)}  I(X; B)_\rho,
\end{gather}
where \(\rho_{XB}=\sum_{x\in\cX} P_X(x)\ketbra{x}\otimes \cW(x) \).
\label{thm:capID.CC}
\end{theorem}

While the single-user achievability proof
in~\cite{ahlswedeDueck1989id1,Loeber1999PhD} employs
a random binning scheme based on transmission codes~\cite{ahlswedeDueck1989id2},
we will see that the broadcast coding methods are significantly more involved
and do not follow from the transmission characterization.

\section{Results}
\label{sec:results}

Our results are presented below.
Consider the quantum broadcast channel
\(\channel_{A\to B_1 B_2}\), as defined in Subsection \ref{subsec:Qchannel}.
Define the rate region $\scrR(\channel)$ as
\begin{gather}
  \scrR(\channel) =
  \bigcup_{
          P_X \in \cP(\cX) ,\; |\phi_A^x\rangle
} \set[\Bigg]{
  \begin{array}{l l}
    (R_1, R_2) :
        & \displaystyle R_1 \leq I(X; B_1)_\rho, \\
   		& \displaystyle R_2 \leq I(X; B_2)_\rho
   	     \end{array}
  }
\end{gather}
with 
$
    \rho_{XB_1 B_2}=\sum_{x\in\cX} P_X(x)\ketbra{x}\otimes\channel(\ketbra{\phi_A^x}).
$

\begin{theorem}
$\,$
\begin{enumerate}
\item
The region $\scrR(\channel)$ is achievable for identification over the quantum broadcast channel $\channel_{A\to B_1 B_2}$.
That is, 
\begin{equation}
  \capSAID(\channel) \supseteq 
  \scrR(\channel).
  \label{eq:thm.capSAID-CBC.bound1}
\end{equation}
\item
The identification capacity region of a classical-quantum broadcast channel $\channel^{\,\text{c-q}}_{X\to B_1 B_2}$ is given by
\begin{gather}
  \capSAID(\channel^{\,\text{c-q}})
  = \bigcup_{ P_X \in \cP(\cX) } \set[\Bigg]{
  \begin{array}{l l}
    (R_1, R_2) :
        & \displaystyle R_1 \leq I(X; B_1)_\rho, \\
   		& \displaystyle R_2 \leq I(X; B_2)_\rho
   	     \end{array}
  },
  \label{eq:thm.capSAID-CBC.bound2}
\end{gather}
with 
$\rho_{XB_1 B_2}=\sum_{x\in\cX} P_X(x)\ketbra{x}\otimes\channel^{\,\text{c-q}}(x)$.
\end{enumerate}

\label{thm:capSAID-CBC}
\end{theorem}
\noindent
The proof of part 1 is given in Section \ref{sec:proof.capSAID-CBC.achiev}, where
we show that all rate pairs in the interior of the region $\scrR(\channel)$
are achievable. In Section \ref{sec:proof.capSAID-CBC.converse}, we prove part 2 and
show the classical-quantum converse part, \ie that no rate pair outside the region above
can be achieved for identification over the classical-quantum broadcast channel. In the proof of part 1, we  use the
pool-selection method by Bracher and Lapidoth~\cite{bracherLapidoth2017idbc,bracher2016PhD}.
This will enable the same extension to the broadcast setting as
in~\cite{bracherLapidoth2017idbc,bracher2016PhD}.
On the other hand, in the converse proof, we used a different approach
exploiting recent observations by
Boche~et~al.~\cite{BocheDeppeWinter2019quantum} along with the methods of
Ahlswede~and~Winter~\cite{AhlswedeWinter2002quantumID}.

\begin{remark}
As mentioned in Subsection~\ref{sec:relatedWork},
in the classical-quantum single-user setting,
the ID and transmission capacity characterizations
are identical. On the other hand, in the broadcast ID setting, we see a departure from this
equivalence \cite{ahlswede2008gtid_updated,bracherLapidoth2017idbc,bracher2016PhD}.
The examples in the following section demonstrate this departure in a more explicit manner,
showing that the ID capacity region can be strictly larger than the
transmission capacity region.
\end{remark}

\begin{remark}
\label{remark:Marginal_Capacity}
Consider the classical-quantum broadcast channel.
In general, the rate $R_k$ of User $k$ must be limited by the ID capacity of the single-user channel from $A$ to $B_k$,
for $k\in\{1,2\}$. 
This observation leads to the following rectangular upper bound,
\begin{align}
  \capSAID(\channel^{\,\text{c-q}})
  \subseteq
  \set[\Bigg]{
    \begin{array}{l l}
      (R_1, R_2) : & R_1 \leq \capID^{(1)}
      , \\
		   & R_2 \leq \capID^{(2)}
    \end{array}
  },
  \label{eq:UpperRect}
\end{align}
where $C_{\text{ID}}^{(k)}
= \max_{P_X}  I(X; B_k)_\rho$.
However, in identification over the broadcast channel, the users cannot necessarily achieve the full capacity of
each marginal channel simultaneously, 
since both marginal channels must share the same input distribution in the capacity formula  on the right hand side of (\ref{eq:thm.capSAID-CBC.bound2}).
Equality holds in (\ref{eq:UpperRect}) if the same input distribution \(P_X^\star\) maximizes
both  mutual informations simultaneously, \ie when
\begin{equation}
  P_X^\star = \argmax_{P_X  }  I(X; B_1)_\rho
  = \argmax_{P_X}  I(X;B_2)_\rho.
  \label{eq:P_X_max_symmetric}
\end{equation}
\end{remark}

\section{Examples}
\label{sec:examples}

As examples, we consider the pure-loss bosonic broadcast channel and the erasure broadcast channel.

\subsection{Bosonic Broadcast Channel}

To demonstrate our results, consider the single-mode bosonic broadcast channel.
We extend the finite-dimension result in Theorem~\ref{thm:capSAID-CBC} to the
bosonic channel with infinite-dimension Hilbert spaces based on the
discretization limiting argument by Guha \etal
\cite{GuhaShapiroErkmen:07p}. A detailed description of
(continuous-variable) bosonic systems can be found in \cite{WPGCRSL:12p}. Here,
we only define the notation for the quantities that we use.
We use hat-notation, \eg $\ha$, $\hb_1$, $\hb_2$, $\he$, to denote
\replaced{
annihilation
operators that act on a
quantum state.}{operators that act on a
  quantum state. The single-mode Hilbert space is spanned by the Fock basis 
  $\{ |n\rangle \}_{n=0}^\infty$. Each $|n\rangle$ is an eigenstate of the number
  operator $\hn=\ha^\dagger \ha$, where $\ha$ is the bosonic field annihilation
  operator. In particular, $|0\rangle$ is the vacuum state of the field. 
  The \emph{creation operator} $\ha^{\dagger}$ creates an excitation: 
  $\ha^{\dagger}|n\rangle=\sqrt{n+1}|n+1\rangle$, for $n\geq 0$. Reversely, the
  \emph{annihilation operator} $\ha$ takes away an excitation:
  $\ha|n+1\rangle=\sqrt{n+1}|n\rangle$.}
A thermal state $\tau(N)$ is a Gaussian mixture of coherent states, where
\(   \tau(N)
  \equiv \int_{\mathbb{C}} d^2 \alpha \frac{e^{-|\alpha|^2/N}}{\pi N}
    |\alpha\rangle \langle \alpha|
   \), with an average photon number $N> 0$. 

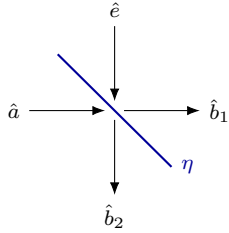
\begin{figure}[tb]
\centering
\begin{tikzpicture}[>=Latex,scale=0.75]
  \node (splitter) {};
  
    \draw[blue!60!black,thick] (-1,1)  -- (1,-1) node[right] {$\eta$};
  
    \draw[thin,<-] (splitter) -- +(-1.5,0) node[left] {$\hat{a}$};
  
    \draw[thin,<-] (splitter) -- +(0,1.5) node[above] {$\hat{e}$};
          
    \draw[thin,->] (splitter)  -- +(1.5,0) node[right] {$\hat{b}_1$};
  
     \draw[thin,->] (splitter)   -- (0, -1.5) node[below] {$\hat{b}_2$};

\end{tikzpicture}
\caption{\label{fig:BSp}
  The beam splitter relation of the single-mode bosonic broadcast
  channel. The channel input is an electromagnetic field mode with
  annihilation operator $\ha$, and the output is a pair of modes with
  annihilation operators $\hb_1$ and $\hb_2$, corresponding to each
  receiver. The mode $\he$ is associated with the environment noise
  in the pure-loss setting, the environment is in the vaccuum state, i.e., $\he = \ket 0$.
  The parameter $\eta$ is the transmissivity,
  which captures the length of the optical fiber and its absorption length.}
\end{figure}

Consider a bosonic broadcast channel, whereby the channel input is an
electromagnetic field mode with  annihilation operator $\ha$, and the output is
a pair of modes with  annihilation operators $\hb_1$ and $\hb_2$.
The annihilation operators correspond to Alice, Bob 1, and Bob 2, respectively. 
The input-output relation of the pure-loss bosonic broadcast channel in the
Heisenberg picture~\cite{HolevoWerner2001bosonic} is given by 
\deleted{bosonic channel is given by}
\begin{align}
  \hb_1 &=\sqrt{\eta}\, \ha +\sqrt{1-\eta}\,\he, \\
  \hb_2 &=\sqrt{1-\eta}\, \ha -\sqrt{\eta}\,\he\added{,}
\end{align}
where $\he$ is associated with the environment noise\deleted{ which is
the vacuum state $|0\rangle$,}
and\deleted{ the parameter} $\eta$ is  the transmissivity, $0\leq \eta\leq 1$,
which captures, for instance, the length of the optical fiber and its absorption
length \cite{EisertWolf2005bosonic}.
The relations above correspond to the outputs of a beam splitter, as illustrated in Figure~\ref{fig:BSp}.
\added{In the pure-loss setting, the environment is in the vacuum state, i.e., $\he = \ket 0$.}
It is assumed that the encoder uses a coherent state protocol with an input constraint.
That is, the input state is a coherent state $|x\rangle$, $x\in\mathbb{C}$,
such that each codeword satisfies
$\frac{1}{n}\sum_{i=1}^n |x_{i}|^2\leq N_{A}$.

Based on part 1 of Theorem~\ref{thm:capSAID-CBC}, the ID capacity region of the
pure-loss bosonic broadcast channel with coherent encoding and average photon number
at most $N_A$
is given by
\begin{align}
 \capSAID(\channel)&= 
\set[\Bigg]{ \begin{array}{rl}
  (R_1,R_2) \,:\;
	R_1 &\leq g(\eta N_{A})    \\
    R_2   &\leq g((1-\eta)N_{A})
	\end{array}
}
\label{eq:inRscG}
\end{align}
where $g(N) = (N+1)\log(N+1)-N\log(N)$
is the entropy of a thermal state with mean photon number $N$,
with $0 \log 0 \coloneqq 0$.
See Figure~\ref{fig:bosonicBC}.
The converse part immediately follows from the
single-user capacity characterization.
To show achievability, set the input to be an ensemble of coherent states,
$|X\rangle$, with a circularly-symmetric Gaussian distribution with zero mean
and variance $\expect[\abs X^2] = N_{A}$.
As mentioned in Remark~\ref{remark:Marginal_Capacity},
the users cannot necessarily achieve the marginal capacity. Nevertheless,
for the bosonic broadcast channel, each user achieves the full capacity of
the respective marginal channel.

On the other hand, the transmission capacity region of the single-mode pure-loss
bosonic broadcast channel is \cite{GuhaShapiro:07c,GuhaShapiroErkmen:07p},
\begin{gather}
\capT(\channel)
  =  \bigcup_{0\leq\beta\leq 1} \set[\Bigg]{ \begin{array}{ll}
  (R_1,R_2) :
    & R_1 \leq g(\eta\beta N_{A}) \\
    & R_2 \leq g((1-\eta) N_{A})-g((1-\eta)\beta N_{A})
	\end{array}
} .
\label{eq:inR0G}
\end{gather}
where the subscript `T' stands for `Transmission', under the assumption that the minimum output-entropy conjecture holds (see Strong Conjecture 2 in \cite{GuhaShapiroErkmen:07p}).
The transmission capacity region and the ID capacity region are depicted in Figure~\ref{fig:bosonicBC}
as the light gray area (T) and additionally the dark gray area (ID), respectively.

The converse part for the transmission result on the pure-loss bosonic broadcast channel relies on the strong minimum output-entropy conjecture \cite{GuhaShapiro:07c}, as stated below. Let the noise modes $\{ \he_i   \}_{i=1}^n$ be in a product state $\rho_{E^n}=\ketbra{0}^{\otimes n}$ of $n$ vacuum states, and assume that $H(A^n)_\rho=n g( N_A)$. Then, the strong minimum output-entropy conjecture states that \cite{GuhaShapiro:07c}
\begin{align}
 H(B^n)_\rho \geq n g (\eta N_A) \,.
\end{align}
We note that in the single-user case, the conjecture is not required for neither identification nor transmission \cite{GGLMSY:04p,QiWildeGuha:16a,DePalmaTrevisanGiovannetti:17p} \cite[Sec. VI.B]{peregFerraraBloch2021key_secrecy_bosonicBC_arxiv}. There are several special cases that are known to hold  \cite[Sec. III]{DePalma:19p}. E.g., it is well known that the conjecture holds for $n=1$.  However, as pointed out in \cite[Sec. V]{DePalmaTrevisanGiovannetti:17p}, this is insufficient for the converse proof of the bosonic broadcast channel, which requires the strong minimum output-entropy conjecture stated above.

\subsection{Erasure Broadcast Channel}

We consider the qubit erasure broadcast channel, specified by
$\channel(\rho)=U\rho U^\dagger$,
\begin{IEEEeqnarray}{l L L}
  U= \sqrt{1-\lambda}\identity_{A\to B_1}\otimes \ket{e}_{B_2}
  +\sqrt{\lambda} \ket{e}_{B_1}\otimes \identity_{A\to B_2} \,,
\end{IEEEeqnarray}
where the erasure state $\ket{e}$ is orthogonal to the qubit space, and $0\leq\lambda\leq \frac 1 2$ is a given parameter.
Hence, the marginal channels to Bob 1 and Bob 2 are standard quantum erasure channels, with erasure parameters $\lambda$ and $1-\lambda$, respectively.
Specifically,
\begin{align}
    \channel^{(1)}(\rho)&=(1-\lambda)\rho+
    \lambda\ketbra{e}\,,\\
    \channel^{(2)}(\rho)&=\lambda\rho+
    (1-\lambda)\ketbra{e} \,.
\end{align}
The ID capacity region of the erasure broadcast channel $\channel_{A\to B_1 B_2}$
satisfies
\begin{align}
  \capSAID(\channel)
  \supseteq
  \scrR(\channel)
  =
  \set[\Bigg]{
    \begin{array}{l l}
      (R_1, R_2) : & R_1 \leq 1-\lambda, \\
		   & R_2 \leq \lambda
    \end{array}
  }\,.
  \label{Equation:Achievable_Erasure}
\end{align}
This result is obtained in a straightforward manner.
To show achievability, we apply part 1 of Theorem~\ref{thm:capSAID-CBC} and set
$P_X=\left(\frac{1}{2},\frac{1}{2}\right)$ over the ensemble $\set{\ket 0, \ket 1}$.

First, consider the symmetric case of $\lambda = \frac 1 2$. Our achievable region  $\scrR(\cN)$ is then the
best known bound 
on the ID capacity region.
Whereas, for $\lambda < \frac 1 2$, we can improve upon our bound.

\begin{figure}[tb]
  \centering
  \providecolor{identification}{named}{blue}
\providecolor{idTimeDivision}{named}{orange}
\providecolor{transmission}{named}{red}

\newcommand{\rateRegionQEBC}[2][]{

\pgfkeys{/pgf/fpu/output format=fixed}
\tikzmath{
 real \e, \Ra, \Rb;
 \e = #2;
 \Ra = Ibec(0.5, \e);
 \Rb = Ibec(0.5, 1-\e);
 function R1ts(\x) { return 2*\x*\Ra; };
 function R2ts(\x) { return (1-\x)*\Rb; };
 function R1t(\x) { return \Ra*\x; };
 function R2t(\x) { return \Rb*(1-\x); };
}
\pgfkeys{/pgf/fpu=false}

\begin{tikzpicture}[auto,thick,#1]
\pgfkeys{/pgf/number format=fixed}
\pgfkeys{/pgf/number format/precision=3}
\draw[->] (0,0) -- (0, {R2ts(0)+0.075}) node[above] {$R_2$};
\draw[->] (0,0) -- ({R1ts(1)+0.075}, 0) node[right] {$R_1$};

\fill[rateRegion/identification] (0,0) -- (0, \Rb) -- +(\Ra, 0) node[below left,color=white] {\large$\scrR (\mathcal \cN)$} -- (\Ra, 0);
\draw (0, \Rb) -- +(-0.03,0) node [left] {$\lambda$};
\draw (\Ra, 0) -- +(0,-0.03) node [below] {$1-\lambda$};
\draw ({R1ts(1)}, 0) -- +(0,-0.03) node [below] {$2 (1-\lambda)$};

\pgfkeys{/pgf/fpu/output format=fixed}
\fill[rateRegion/idTimeDivision] (0,0) -- plot [smooth,parametric,domain=0:1,range=0:1,samples=100] ({R1ts(\x)}, {R2ts(\x)});
\pgfkeys{/pgf/fpu=false}
\draw[color=white] ({R1ts(0.5)}, {R2ts(0.5)}) node[below,yshift=-0.5em] {\large$\mathscr T$};

\fill[rateRegion/transmission] (0,0) -- plot [smooth,parametric,domain=0:1,range=0:1,samples=100] ({R1t(\x)}, {R2t(\x)});
\draw ({R1t(0.5)}, {R2t(0.5)}) node[below left] {\large$\capT(\mathcal N)$};
\end{tikzpicture}
}
  \def\lambdaTikz{0.4}
  \rateRegionQEBC[scale=5]{\lambdaTikz}
  \caption[ID and transmission capacity regions of the bosonic broadcast channel]{     \label{fig:qebc}     Achievable regions
     for ID over the qubit erasure broadcast channel $\channel$ with
     erasure probability $\lambda = \frac 1 4$.
     The transmission capacity region $\capT(\cN)$ corresponds to the 
     light gray area. The region $\scrT$ achievable by time-division between
     single-user identification codes comprises additionally the middle gray area.
     The rectangle indicated by the dark gray area corresponds
     to our lower bound $\scrR(\cN)$.}
\end{figure}
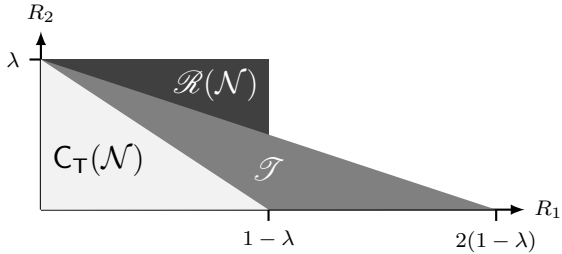

For a single-user quantum erasure channel $\mathcal{L}_{A\to B}$ with a parameter $\varepsilon$, Winter established achievability of the identification rate
$R = 2(1-\varepsilon)$
for $\varepsilon< \frac 1 2$,
and $R = 1-\varepsilon$ for $\varepsilon\geq \frac 1 2$ \cite[Section~4]{Winter:13b}, where $\rho_X$ is a classical state.
If $\lambda = \frac 1 2$,
then this yields the rate pair
$(R_1,R_2)=\left( \frac 1 2, \frac 1 2 \right)$, which is the corner of our region
in~\eqref{Equation:Achievable_Erasure}.
On the other hand, for $\lambda < \frac 1 2$, 
the rate pairs
$(R_1,R_2) = (2(1-\lambda),0)$ and
$(R_1,R_2)=(0,\lambda)$ are achievable.
Hence, by time division, i.e.
coding for User 1 over a sub-block of length $\alpha n$, and for User 2 over the remaining sub-block of $(1-\alpha)n$ channel uses,
we have that 
the ID capacity region is lower-bounded by
\begin{align}
  \capSAID(\channel)
  \supseteq
  \scrT
  \coloneqq
  \bigcup_{0 \le \alpha \le 1}
  \set[\Bigg]{
    \begin{array}{l l}
      (R_1, R_2) : & R_1 \le 2 \alpha (1-\lambda), \\
                   & R_2 \le (1-\alpha)\lambda
    \end{array}
  }\,.
\end{align}
The transmission capacity region $\capT(\channel)$
of the quantum erasure broadcast channel is also achieved by time-division.
It is given by the rate pairs $(R_1, R_2)$ satisfying
$R_1 \le \alpha (1-\lambda)$ and $R_2 \le (1-\alpha)\lambda$,
for some $0 \le \alpha \le 1$,
as can be shown using the same methods used in~\cite[Example~3.2 and Section~5.4.1]{elgamalKim2011network_it} 
and~\cite[Section~20.4.3]{wilde2017quantum_it_2}.
Clearly, $\capT(\channel)$ is contained in either of the regions $\scrR(\channel)$, $\scrT$.
We deduce that for an erasure channel that is not symmetric, our achievable
region $\scrR(\channel)$ in {Theorem~\ref{thm:capSAID-CBC}}
is suboptimal, but improves on the best previously known bound
in the interval $0 < R_1 \le 1 - \lambda$.
Figure~\ref{fig:qebc} shows the regions $\capT(\channel)$, $\scrT$ and $\scrR(\channel)$
for a quantum erasure broadcast channel $\channel$ with $\lambda = \frac 1 4$.

\section{Achievability Proof}
\label{sec:proof.capSAID-CBC.achiev}

In this section we \replaced{prove the lower bound on}{lower-bound} the capacity region in Theorem \ref{thm:capSAID-CBC}, \ie we show that
\begin{equation}
  \capSAID(\channel) \supseteq \scrR(\channel).
\end{equation}

In the classical achievability proof, Bracher and Lapdidoth
\cite{bracherLapidoth2017idbc,bracher2016PhD} 
first generate a single-user random code, based on a
pool-selection technique, as shown below. Then, a similar
pool-selection code is constructed for the BC using a
pair of single-user codes, one for each receiver. It is shown in
\cite{bracherLapidoth2017idbc,bracher2016PhD} that the
corresponding ID error probabilities for the BC can be
approximated in terms of the error probabilities of the single-user
codes. We use a similar approach, and begin with the \emph{single-user}
quantum channel.

We use standard tools of typical space projectors as detailed in Appendix~\ref{sec:quantum_types}.
In particular, \(\cT_\delta^n(P_X)\) denotes  the classical \(\delta\)-typical set with respect to a given PMF  \(P_X \in \cP(\cX)\)  over \(\cX\). Furthermore, \(\Pi_\delta^n(\rho)\) is the projector onto the $\delta$-typical subspace of an average state $\rho = \sum_{x \in \cX} P_X(x) \ketbra{x}$, and 
\(\Pi_\delta^n(\sigma_{XB}|x^n)\) is the conditionally $\delta$-typical projector for a classical-quantum state \(\sigma_{XB}\).

\subsection{Single-User Quantum Channel}
\label{sec:proof.capSAID-CBC.achiev.single}

\added{First, we construct and analyze an identification code for a single-user quantum
channel. In Section~\ref{sec:proof.capSAID-CBC.achiev.broadcast}, we will use
the single-user code in order to construct a code for the broadcast channel.}
Let \(\cL_{A\to B}\) be a single-user quantum channel.
\subsubsection{Code Construction}

Let \(N = \exp e^{nR}\) be the code size.
Fix  a PMF $P_X$ over \(\cX\),
a pool rate \(\Rpool\), and a binning rate \(\Rbin\), such that
\begin{align}
  R      &< \Rbin < I(X;B)_\rho \\
  \Rpool &> \Rbin.
\end{align}
We generate the codebook such that all codewords are $\delta$-typical.
Therefore, consider
the distribution
\begin{gather}
  P_{X'^n}(x^n) = \frac{P_X^n(x^n)}{P_X^n \tup{\cT_\delta^n(P_X)} } \cdot \ind{ x^n \in \cT_\delta^n(P_X) },
\end{gather}
where the indicator function $\ind{\pi}$ takes the value 1 if $\pi$ is true,
and 0 otherwise.
For every index \(v \in \cV = [e^{n\Rpool}]\),
choose a codeword \(\Xpool(v) \sim P_{X'^n}\) at random. Then, for every
\(i \in [N]\), decide whether to add \(v\) to the set \(\bcV_i\) by a binary
experiment, with probability
\(e^{-n\Rbin}/\abs{\cV} = e^{-n(\Rpool - \Rbin)}\). That is, decide to
include \(v\) in \(\bcV_i\) with probability \(e^{-n(\Rpool - \Rbin)}\),
and not to include it with probability \(1-e^{-n(\Rpool - \Rbin)}\). Reveal
this construction to all parties.
Denote the collection of codewords and index bins by
\begin{equation}
  \CodeBL = \tup[\Big]{\{F(v)\}_{v\in\cV}, \set{\bcV_i}_{i=1}^N} .
\end{equation}

\subsubsection{Encoding}

To send an ID Message \(i \in [N]\), Alice chooses an index \(v\)
uniformly at random from \(\bcV_i\).
If \(\bcV_i\) is non-empty, she prepares the state 
\begin{align}
    \ket{ \phi_{A^n}^{F(v)} } \equiv \bigotimes_{t=1}^n \ket{ \phi_A^{F_t(v)} },
\end{align}
where $F_t(v)$ is the $t$-th symbol of the sequence $F(v)$.
Otherwise, if \(\bcV_i=\emptyset\),  she prepares $|\phi_{A^n}^{F(1)}\rangle$.
Then, Alice transmits the systems $A^n$ through the channel.
Therefore, if $\bcV_i \neq \emptyset$, then the average input state \(\cE_{A^n}(i)\) is given by
\begin{equation}
  \cE_{A^n}(i) =     \frac{1}{\abs{\bcV_i}} \sum_{v \in \bcV_i} 
      \ketbra{\phi_{A^n}^{F(v)}} .
  \label{eq:capSAID.proof.enc.single-user}
\end{equation}

\subsubsection{Decoding}
Bob receives the output systems $B^n$ and he would like to determine whether the
message $i'$ was sent. To this end,
he selects any constant $\delta$ such that
\begin{gather}
  \label{eq:deltaBound}
  0 < \delta < \frac{I(X; B)_\rho - \Rbin}{c+c'},
\end{gather}
where $c, c' > 0$ are constants as in Section \ref{sec:quantum_types}.
Then, he performs a series of binary decoding measurements (POVMs)
\begin{gather}
  \cD^{F(v)} = \set{ \identity - \Pi \Pi^{F(v)} \Pi, \Pi \Pi^{F(v)} \Pi},
\end{gather}
where we denote $\Pi \equiv \Pi_\delta^n(\rho_B)$ and $\Pi^{F(v)} \equiv \Pi_\delta^n(\rho_{XB}|F(v))$.
Bob obtains a binary sequence of measurement outcomes $(a(v))_{v \in \bcV_{i'}}$.
If $a(v)=1$ for some $v\in\bcV_{i'}$, then Bob declares that $i'$ was sent.
Otherwise, he declares that \(i'\) was not sent. Note that we can also construct one
POVM $\bcD_{B^n}^{i'}$ that is equivalent to the series of measurements.

Thus, the ID code associated with the construction above is denoted by
\[\cC_{\CodeBL} = \tup{\bcE_{A^n}, \bcD_{B^n}}.\]
The error analysis for the single-user identification code is delegated to
Appendix~\ref{sec:achiev.single.error_analysis}.

\subsection{Broadcast Channel}
\label{sec:proof.capSAID-CBC.achiev.broadcast}

In this section, we show the direct part for the ID capacity
region of the quantum broadcast channel. That is, we show that \(\capSAID(\channel) \supseteq \scrR(\channel)\).
The analysis makes use of the our single-user derivation above.

\subsubsection{Code Construction}

We extend Bracher and Lapdioth's \cite{bracherLapidoth2017idbc,bracher2016PhD}
idea to combine two BL codebooks \(\CodeBL^{(1)}, \CodeBL^{(1)}\) that share the same pool.
Fix a PMF \(P_X\) over \(\cX\) and rates \(R_k,\Rbin_k\), for \(k\in\{1,2\}\), that satisfy
\begin{subequations} \label{eq:capSAID.CBC.achiev.constr.rates}
\begin{IEEEeqnarray}{rCL}
  R_1 < &\Rbin_1        &< \min_{s \in \cS} I(X; B_1)_\rho \\
  R_2 < &\Rbin_2        &< \min_{s \in \cS} I(X; B_2)_\rho \\
	& \max\set[\big]{\Rbin_1, \Rbin_2} &< \Rpool \\
  \Rpool < & \Rbin_1 + \Rbin_2.
\end{IEEEeqnarray}
\end{subequations}
Let $N_k = e^{nR_k}$.
For every index \(v \in \cV = [e^{n\Rpool}]\), perform the following. Choose
a codeword \(F(v) \sim P^n_X\) at random, as in the single-user case.
Then, for every \(i_k\), decide whether to add \(v\) to the set \(\bcV_{i_k}^{(k)}\) by a binary
experiment, with probability
\(e^{-n\Rbin_k}/\abs{\cV} = e^{-n(\Rpool - \Rbin_k)}\). That is, decide that
\(v\) is included in \(\bcV_{i_k}^{(k)}\) with probability \(e^{-n(\Rpool - \Rbin_k)}\),
and not to include with probability \(1-e^{-n(\Rpool - \Rbin_k)}\). 
Finally, for every pair \((i_1,i_2) \in [N_1] \times [N_2]\), select a common index \(V_{i_1,i_2}\)
uniformly at random from \(\bcV_{i_1}^{(1)} \cap \bcV_{i_2}^{(2)}\), if this intersection is non-empty.
Otherwise, if \(\bcV_{i_1}^{(1)} \cap \bcV_{i_2}^{(2)} = \emptyset\), then draw \(V_{i_1,i_2}\) uniformly from \(\cV\).
Reveal this construction to all parties.

Denote the collection of codewords and index bins by \begin{align}
  \CodeBL_{\channel} = \tup[\Big]{\Xpool
    ,\, \set[\big]{\bcV_{i_1}^{(1)}}_{i_1 \in [N_1]}&, \set[\big]{\bcV_{i_2}^{(2)}}_{i_2 \in [N_2]}
    ,\, \set[\big]{V_{i_1,i_2}}_{(i_1,i_2) \in [N_1] \times [N_2]}
    }.
\end{align}
Note that, for $k \in \set{1,2}$, \(\CodeBL_\channel\) includes all elements of
$
\CodeBL^{(k)} = \tup[\big]{ \Xpool, \set[]{\bcV_{i_k}^{(k)}}_{i_k \in [N_k]} },
  $
defined for the marginal channels $\channel^{(k)}_{A\to B_k}$
as in Section \ref{sec:proof.capSAID-CBC.achiev.single}.
We denote the corresponding single-user code by
\begin{gather}
  \codet_{\CodeBL^{(k)}} = (\btcE_{A^n}^{(k)}, \bcD_{B_k^n}).
\end{gather}

\subsubsection{Encoding}

To send an ID message pair \((i_1,i_2) \in [N_1] \times [N_2]\),
Alice prepares the input state $\ket[\big]{ \phi_{A^n}^{F(V_{i_1,i_2})} }$
and transmits the input system $A^n$.

\subsubsection{Decoding}

Receiver \(k\), for \(k=1,2\), employs the decoder of the
single-user code \(\codet_{\CodeBL^{(k)}}\).
Specifically, suppose that Bob $k$ is interested in an ID message \(i'_k \in [N_k]\).
Then, he uses the decoding POVM \(\bcD_{B_k^n}^{i'_k}\) to decide whether \(i'_k\) was sent or not.

We denote the broadcast ID code associated with the construction above by
\begin{equation}
  \codet_{\CodeBL_\channel} = (\bcE_{A^n}, \bcD_{B^n_1}, \bcD_{B^n_2})
\end{equation}

\subsubsection{Error Analysis}

We show that the semi-average error probabilities of the ID code
defined above can be approximately upper-bounded by the respective error probabilities
of the single-user ID-codes \(\codet_{\CodeBL^{(1)}}\) and \(\codet_{\CodeBL^{(2)}}\) for the
respective receivers.

Consider a given pair of codebooks $\CodeBL^{(1)} $ and $\CodeBL^{(2)} $.
Conditioned on those codebooks, the input state can be written in terms of an encoding distribution \begin{align}
    \bcE_{A^n}(i_1,i_2) = \sum_{v \in \cV}
    \Enc[i_1,i_2](v) \ketbra{\phi_{A^n}^{F(v)}} \,,
    \label{eq:BinState}
\end{align}
where $\Enc[i_1,i_2](v) = \ind{v = V_{i_1,i_2}}$.
Similarly,
\begin{align}
    \btcE^{(k)}_{A^n}(i_k) = \sum_{v \in \cV}
    \EncAlt[i_k]^{(k)}(v)\ketbra{\phi_{A^n}^{F(v)}}
    \label{eq:BinState1}
\end{align}
where $\EncAlt[i_k]^{(k)}(v)$ is the respective distribution for the single-user
code from Section~\ref{sec:proof.capSAID-CBC.achiev.single}, namely
\begin{gather}
  \EncAlt[i_k]^{(k)}(v)
  = \begin{cases}
    \frac{1}{\abs[\big]{\bcV_{i_k}^{(k)}}} \ind{v \in \bcV_{i_k}^{(k)}} & \text{if $\bcV_{i_k}^{(k)} \neq \emptyset$}\,, \\
    \ind{v = 1} & \text{if $\bcV_{i_k}^{(k)} = \emptyset$}\,.
  \end{cases}
  \label{eq:capSAID.proof.enc.cq.single-user}
\end{gather}

We consider now only Receiver 1 and his
marginal channel \(\channel_{A_1\to B}^{(1)}\).
Since the code construction is completely symmetric between the
two receivers, the same arguments hold for Receiver 2 and  \(\channel_{A_2\to B}^{(2)}\).
The missed-ID error probability for $\codet_{\CodeBL_\channel}$ and Bob 1 is given by
\begin{align}
  \saerrya
    (\channel, n, \codet_{\CodeBL_\channel}, i_1)
    = \frac{1}{N_2} \sum_{i_2 \in [N_2]}
      \trace\intv[\Big]{
        (\identity-D_{i_1}^{(1)})\channel^{(1)\otimes n}_{A \to B_1}\tup{ \bcE_{A^n}(i_1,i_2) }
      }
\end{align}
and for $\codet_{\CodeBL^{(k)}}$ it is given by
\begin{align}
  \erra
    (\channel^{(1)}_{A \to B_1}, n, \codet_{\CodeBL^{(1)}}, i_1)
    =
      \trace\intv[\Big]{
        (\identity-D_{i_1}^{(1)})\channel^{(1)\otimes n}_{A \to B_1}\tup{ \btcE_{A^n}^{(k)}(i_1) }
      }.
\end{align}
By the linearity of the channel and the measurement, we have
\begin{align}
  &\saerrya (\channel, n, \codet_{\CodeBL_\channel}, i_1)
  - \erra (\channel^{(1)}_{A \to B_1}, n, \codet_{\CodeBL^{(1)}}, i_1)
  \nonumber\\ &
    \leq \frac{1}{2} \norm{
      \frac{1}{N_2} \sum_{i_2 \in [N_2]} \bcE_{A^n}(i_1 ,i_2)
      - \btcE_{A^n}^{(k)}(i_1)
    }_1
  \nonumber\\ &
    = \frac{1}{2} \norm{
      \frac{1}{N_2} \sum_{i_2 \in [N_2]} \sum_{v \in \cV} \Enc[i_1,i_2](v) \ketbra{\phi_{A^n}^{F(v)}}
      -  \sum_{v \in \cV} \EncAlt[i_1]^{(k)}(v) \ketbra{\phi_{A^n}^{F(v)}}
    }_1
  \nonumber\\ &
    = \frac{1}{2} \abs{
      \sum_{v \in \cV} \tup{
        \frac{1}{N_2} \sum_{i_2 \in [N_2]} \Enc[i_1,i_2](v)
      -  \sum_{v \in \cV} \EncAlt[i_1]^{(k)}(v)
      }
    }
  \nonumber\\ &
    \leq \delta_{i_1}^{(1)},
\end{align}
where $\delta_{i_1}^{(1)}$ is the total variation distance
\begin{align}
  \delta_{i_1}^{(1)}
    = \frac{1}{2} \sum_{v \in \cV} \abs{
      \frac{1}{N_2} \sum_{i_2 \in [N_2]} \Enc[i_1,i_2](v)
      -  \sum_{v \in \cV} \EncAlt[i_1]^{(1)}(v)
    }
     = d\tup{ \frac{1}{N_2} \sum_{i_2 \in [N_2]} \Enc[i_1,i_2],\, \EncAlt[i_1]^{(1)} },
\end{align}
and the inequalities follow from the triangle inequality.
The same argument applies to the false-ID error.
Hence,
\begin{subequations}
\label{eq:BL-code.err_reduces}
\begin{align}
  \saerrya(\channel, n, \codet_{\CodeBL_\channel}, i_1)
     &\leq \erra(\channel^{(1)}_{A \to B_1}, n, \codet_{\CodeBL^{(1)}}, i_1) +  \delta_{i_1}^{(1)},
     \\
  \saerryb(\channel, n, \codet_{\CodeBL_\channel}, i'_1, i_1)
     &\leq \errb(\channel^{(1)}_{A \to B_1}, n, \codet_{\CodeBL^{(1)}}, i'_1, i_1) + \delta_{i_1}^{(1)},
\end{align}
Similarly, the error probabilities for the second marginal channel are bounded by
\begin{align}
  \saerrza(\channel, n, \codet_{\CodeBL_\channel}, i_2)
     &\leq \erra(\channel^{(2)}_{A \to B_2}, n, \codet_{\CodeBL^{(2)}}, i_2) + \delta_{i_2}^{(2)},
     \\
  \saerrzb(\channel, n, \codet_{\CodeBL_\channel}, i'_2, i_2)
     &\leq \errb(\channel^{(2)}_{A \to B_2}, n, \codet_{\CodeBL^{(2)}}, i'_2, i_2) + \delta_{i_2}^{(2)},
\end{align}
where $\delta_{i_2}^{(2)} = d\tup[\Big]{ \frac{1}{N_1} \sum_{i_1 \in [N_1]} \Enc[i_1,i_2],\, \EncAlt[i_2]^{(2)} }$.
\end{subequations}

From this point, we can continue as in the classical derivation due to Bracher and Lapidoth \cite{bracherLapidoth2017idbc,bracher2016PhD}.
The next lemma bounds $\delta_{i_k}^{(k)}$ in~\eqref{eq:BL-code.err_reduces}
to zero in probability as \(n \to \infty\).
By \cite{bracherLapidoth2017idbc}, \cite[Lemma~3]{rosenbergerPeregDeppe2023id_compound_BC},
for every $k \in \set{1,2}$ and some \(\tau > 0\),
\begin{gather}
        \lim_{n\to\infty} \Pr\tup[\bigg]{
    \max_{i_k \in [N_k]} \delta_{i_k}^{(k)} \geq e^{-n\tau}
  } = 0.
\end{gather}

Hence by \eqref{eq:BL-code.err_reduces}, the error probabilities for the quantum broadcast-channel code \(\codet_{\CodeBL_\channel}\)
are approximately upper-bounded by the corresponding error probabilities
for the single-user marginal codes \(\codet_{\CodeBL^{(1)}}\) and \( \codet_{\CodeBL^{(2)}}\).

By \eqref{eq:capSAID-CBC.direct.single-user} for the single-user
quantum channel $\cN^{(k)}_{A \to B_k}$ and $k \in \set{1,2}$, the error probabilities 
  \(\erra(\channel^{(k)}_{A \to B_k}, n, \codet_{\CodeBL^{(k)}}, i_k)\) and
  \(\errb(\channel^{(k)}_{A \to B_k}, n, \codet_{\CodeBL^{(k)}}, i'_k, i_k)\)
converge in probability to zero with convergence speed exponentially in \(n\),
for all messages \(i_k, i'_k \in [N_k]\) such that \(i_k \neq i'_k\).
This completes the proof of the direct part.
\pushQED

\section{Converse Proof}
\label{sec:proof.capSAID-CBC.converse}
The direct part follows from part 1. Hence, it remains to prove the converse part.
To this end,
consider an \((N_1, N_2, n, \lambda_1,\lambda_2)\) ID code,
\(\codet = (\cE_{X^n}, \cD_{B_1^n}, \cD_{B_2^n})\),
for the c-q broadcast channel \(\channel_{X\to B_1 B_2}\).
In the case of a classical input, the encoder effectively assigns a probability distribution $Q_{i_1,i_2}$ to each message pair, \ie
\begin{align}
    \cE_{X^n}(i_1,i_2)=\sum_{x^n\in\cX^n} Q_{i_1,i_2}(x^n)\ketbra{x^n}.
\end{align}
Thus, the ID code is specified
by $\set[\big]{\tup[\big]{Q_{i_1,i_2},  D^{(1)}_{i_1}, D^{(2)}_{i_2} } \,:\; (i_1,i_2)\in [N_1]\times [N_2] }$.
We denote the Holevo information for each c-q channel \(\channel^{(k)}_{X\to B_k}\) with respect to an input distribution $P_X\in\cP(\cX)$ by
    $
    \mathsf{I}(P_X,\channel^{(k)}_{X\to B_k})\equiv I(X;B_k)_\rho .
    $

Following the approach of Boche et al. \cite{BocheDeppeWinter2019quantum}, we prove the converse part in three stages, beginning with a modification of the code.

\subsection{Code Modification}

\subsubsection{\texorpdfstring{$\delta$}{Delta}-net on \texorpdfstring{$\cP(\cX)$}{P(X)}}

First, we fix a $\delta$-net $\cT$ of probability distributions on $\cX$. That is, for $|\cT|\leq \left(\frac{c}{\delta}\right)^{|\cX|}$, there exists $\cT\subseteq\cP(\cX)$ such that
$
    \cX^n=\bigcup_{P\in\cT} \cA_{P}
    $
and such that the type of an input sequence $x^n\in \cA_{P}$ is $\delta$-close to $P$.
Hence,
\begin{align}
   Q_{i_1,i_2}=\bigoplus_{P\in\cT} \mu_{i_1,i_2}(P) Q^P_{i_1,i_2} 
\end{align}
where $\mu_{i_1,i_2}$ is a PMF over $\cT$, and $Q^P_{i_1,i_2}$ are PMFs over $\cA_P$.  

\subsubsection{\texorpdfstring{$\epsilon$}{Epsilon}-net on \texorpdfstring{$\cP(\cT)$}{P(T)}}
For $|\cM|\leq \left(\frac{c}{\epsilon}\right)^{|\cT|}$, there exists an $\epsilon$-net $\cM\subseteq\mathcal{P}(\cT)$.
Hence, for every $i_2$, there exists a PMF $\mu'_{i_2}\in\cM$ such that
at least a fraction $\frac{1}{|\cM|}$ of the messages $i_1\in [N_1]$ has
$\mu_{i_1,i_2}$ that is $\epsilon$-close to $\mu_{i_2}'$.  
Without loss of generality, for $N_1'=\left\lfloor \frac{N_1}{|\cM|}  \right\rfloor$,
\begin{align}
    \forall i_1\in [N_1'] \,:\; \frac{1}{2}\left\lVert \mu_{i_1,i_2}-\mu'_{i_2} \right\rVert_1\leq \epsilon.
\end{align}
Similarly, there exists  a probability distribution $\mu''\in\cM$ such that at least a $\frac{1}{|\cM|}$ of the messages $i_2\in [N_2]$ has $\mu'_{i_2}$ that is $\epsilon$-close to $\mu''$.  Without loss of generality, for $N_2'=\left\lfloor \frac{N_2}{|\cM|}  \right\rfloor$,
$
    \forall i_2\in  [N_2'] \,:\; \frac{1}{2}\left\lVert \mu_{i_2}'-\mu'' \right\rVert_1\leq \epsilon.
$
Thereby,
\begin{align}
    \forall (i_1,i_2)\in  [N_1']\times [N_2'] \,:\; \frac{1}{2}\left\lVert \mu_{i_1,i_2}'-\mu'' \right\rVert_1\leq 2\epsilon.
\end{align}
Then, we modify the encoding distribution and define
\begin{align}
   Q_{i_1,i_2}''=\bigoplus_{P\in\cT} \mu''(P) Q^P_{i_1,i_2} \,,
\end{align}
leaving the decoder as it is. This results in an
$(n,N_1',N_2',\lambda_1+2\epsilon,\lambda_2+2\epsilon)$ code, where we choose $\epsilon$ to be sufficiently small such that $\lambda_1+\lambda_2+2\epsilon<1$.

\subsection{Encoder Truncation}

There exists $P^*\in\cT$ such that $\mu''(P^*)\geq \frac{1}{|\cT|}$. Thereby, we
modify the code once more and truncate all the other distributions in $\cT$.
That is, we consider the code
$\set[\big]{ (Q_{i_1,i_2}^{P^*}, D^{(1)}_{i_1},D^{(2)}_{i_2}) : (i_1,i_2) \in [N_1'] \times [N_2']}$.
For the new code, the error probabilities of the first and the second kind are bounded by 
$\lambda_k^*=\abs\cT(\lambda_k+2\epsilon)$ for 
$k\in\{1,2\}$. 
Letting $\epsilon\equiv \epsilon(\lambda_1,\lambda_2)\to 0$ as $\lambda_1,\lambda_2\to 0$, the error probabilities of the truncated code tend to zero for every given $\delta>0$.

\subsection{Rate bounds}

Consider the marginal $\channel^{(1)}_{A\to B_1}$ and $Q_{i_1}^{P^*} \equiv \frac{1}{N_2} \sum_{i_2=1}^{N_2'}Q_{i_1,i_2}^{P^*}$.
Let $i_2$ be uniformly distributed. Then, observe that the randomized-encoder code
$\set[\big]{ (Q_{i_1}^{P^*}, D^{(1)}_{i_1}) : i_1 \in [N_1'] }$
is an $(n,N_1',\lambda_1^*,\lambda_2^*)$ ID code for the single-user channel $\channel^{(1)}_{A\to B_1}$.
Therefore, following the single-user converse proof by
Ahlswede~and~Winter~\cite{AhlswedeWinter2002quantumID} (see also~\cite[Section III]{BocheDeppeWinter2019quantum}), 
\begin{align}
  R_1= \frac{1}{n} \log\log(N_1')
    &
  <
  \mathsf{I}(P^*,\channel^{(1)}_{X\to B_1})+\epsilon_{1}
  \label{eq:ConvBR1}
    =
  I(X;B_1)_\rho + \epsilon_1
\end{align}
where $X\sim P^*$ and $\epsilon_1$ tends to zero as $\delta \to 0$.
For completeness, we prove the inequality in the appendix.
Similarly, we also have
$
  R_2  <
    I(X;B_2)_\rho+\epsilon_{2} .
$
This completes the proof of the ID capacity theorem.
\pushQED

\section{Summary and Outlook}
\label{sec:summary}

We derive an achievable ID region for  the quantum broadcast channel and established full characterization for the classical-quantum broadcst channel.
To prove achievability, we extende the classical proof due to Bracher and Lapidoth
\cite{bracherLapidoth2017idbc,bracher2016PhD} to the quantum setting.
On the other hand, in the converse proof, we use the truncation approach  by Boche et al. \cite{BocheDeppeWinter2019quantum} along with the arguments of Ahlswede and Winter \cite{AhlswedeWinter2002quantumID}.

As examples,
we derive explicit expressions for the ID capacity
regions for the quantum erasure broadcast channel and for the pure-loss bosonic broadcast channel in Section \ref{sec:examples}.
In those examples, each user can achieve the capacity of the
respective marginal channel. In particular, the ID capacity region of the pure-loss bosonic broadcast channel is rectangular
and strictly larger than the transmission capacity region.
In general, the ID capacity region is not necessarily rectangular, as demonstrated for
the classical Z-channel \cite[Section IV.C]{rosenbergerPeregDeppe2023id_compound_BC}
and the classical Gaussian Product channel \cite[Section IV.E]{rosenbergerPeregDeppe2023id_compound_BC},
\cite[Section IV.B]{rosenbergerPeregDeppe2022id_compound_BC_conference}.

The  ID capacity has a different behavior
compared to the single-user setting, in which the ID capacity equals the
transmission capacity \cite{watanabe2022idMinimaxConverse}  (see Section \ref{sec:relatedWork}).
Here, in the broadcast setting, the ID capacity region can strictly larger than in transmission,
since interference between receivers can be seen as part of
the randomization of the coding scheme. 

Extending the results to more than two receivers remains an
open challenge. Upper and lower bounds for such a model may be derived
in a similar manner as in the classical setting \cite[Section IV.A]{bracherLapidoth2017idbc}.
To derive the identification capacity of classical-quantum channels, new methods are required.
The capacity of quantum-quantum channels is even unknown
for general point-to-point discrete memoryless channels~\cite{Winter:13b,atifPradhanWinter2023quantum_softCover_identification_arxiv}.
These are interesting and challenging directions of further research.

\onecolumn\newpage
\appendix

\section{Quantum Method of Types}
\label{sec:quantum_types}

We review the basic method-of-types properties that will be useful in the analysis.
The \(n\)-type  \(\type{P}_{x^n}\) of a sequence \(x^n \in \cX^n\)
is defined by \(\type{P}_{x^n}(a) = \frac{n(a|x^n)}{n} \) for $a\in\cX$, where $n(a|x^n)$ is the number of occurrences of the letter $a$ in the sequence $x^n$.
The set of all \(n\)-types over a set \(\cX\) is denoted by
\(\cP(n,\cX)\).
Joint and conditional types are defined similarly, as in \cite{elgamalKim2011network_it}.
Furthermore, a \(\delta\)-typical set is defined as follows.
Given a PMF  \(P_X \in \cP(\cX)\)  over \(\cX\), define
the robustly\footnote{This is the similar to strong $\epsilon$-typicality
\cite{wilde2017quantum_it_2}, but we have $\epsilon = \delta P_X(a)$,
which depends on $a$.}
\(\delta\)-typical set,
\begin{gather}
  \cT_\delta^n(P_X) = \set[\big]{
    x^n \in \cX^n :
    \abs[\big]{\type{P}_{x^n}(a) - P_X(a)} \leq \delta \cdot P_X(a),~a \in \cX
  }.
\end{gather}
Given $P_{Y|X} : \cX \to \cP(\cY)$,
the \emph{conditionally} $\delta$-typical set $\cT_\delta^n (x^n)$ is defined as
the set of all sequences $y^n$ such that $(x^n, y^n) \in \cT_\delta^n(\type{P}_{x^n} \times P_{Y|X})$.
We will use the property
(see \cite[Lemma 2.12]{csiszarKoerner2011IT})
\begin{gather}
  \label{eq:probTypicalLB}
  \Pr\tup{ X^n \in \cT_\delta^n(P_X) } \geq 1 - 2 \abs\cX 2^{-2 n \delta^2}
  \,.
\end{gather}

Consider the state $A^n$ of a quantum system generated from an ensemble
$\set{ P_X(x), \ket{x} }_{x \in \Xset}$. Then, the average density operator is
$\rho = \sum_{x \in \cX} P_X(x) \ketbra{x}$.
The projector onto the $\delta$-typical subspace is defined as
\begin{align}
  \Pi_\delta^n(\rho) = \Pi_\delta^{A^n}(\rho) = \sum_{x^n\in\cT_\delta^n(P_X)} \ketbra{ x^n }_{A^n} .
\end{align}
 For every $\delta>0$ and sufficiently large $n$, the $\delta$-typical projector satisfies
\begin{align}
  \trace( \Pi_\delta^n(\rho) \rho^{\otimes n} )&\geq 1-2^{-b\delta n} , \label{eq:UnitT} \\
  2^{-n(H(\rho)+c\delta)} \Pi_\delta^n(\rho) 
    &\preceq \Pi_\delta^n(\rho) \,\rho^{\otimes n}\, \Pi_\delta^n(\rho)
  \nonumber\\
  &\preceq 2^{-n(H(\rho)-c\delta)} \Pi_\delta^n(\rho) ,
\label{eq:rhonProjIneq}
\\
\trace( \Pi_\delta^n(\rho))&\leq 2^{n(H(\rho)+c\delta)} \label{eq:Pidim}
\end{align}
where $b,c>0$ are constants \cite[Subsection 15.1.3]{Wilde:17b},
while the exponential convergence in (\ref{eq:UnitT}) follows from
Hoeffding's inequality~\cite[Theorem~1]{hoeffding1963inequalities}
(see proof of \eqref{eq:probTypicalLB}
in~\cite[Lemma 2.12]{csiszarKoerner2011IT}).

We will also need conditionally $\delta$-typical subspaces.
Consider a joint classical-quantum system $(X^n, B^n)$ with
density matrix $\sigma_{XB}^{\otimes n}$.
Then, let $B_a(x^n) = \bigotimes_{i : x_i = a} B_i$ be 
the subsystem of $B^n$ with indices $i$ such that $x_i = a$,
and note that there exists a cq-channel
\replaced{$\cN_{X \to B} : a \mapsto
  \tup{\bra{a}\otimes\identity} \sigma_{XB} \tup{\ket{a}\otimes\identity}
  / P_X(a)$.}{$\cN_{X \to B} : a \mapsto \bra{a}\sigma_{XB}\ket{a} / P_X(a)$.}
As in \cite[Definition 15.2.3]{Wilde:17b}, the conditionally
$\delta$-typical projector $\Pi_\delta^n(\sigma_{XB}|x^n)$ is defined by
\begin{gather}
  \Pi_\delta^n(\sigma_{XB}|x^n)
  = \bigotimes_{a \in \cX}
  \Pi_\delta^{t(a|x^n)} \tup{ \cN_{X \to B}(a) },
\end{gather}
where $t(a|x^n) \equiv \set{ t : x_t = a }$, and $\Pi_\delta^{t(a|x^n)}(\sigma_B)$
is the projector onto the subspace of $\cH_B^{\otimes n}$
where the positions in $t(a|x^n)$ are $\delta$-typical for $\sigma_B$.
Given $x^n \in \cT_\delta^n(P_X)$,
it satisfies
\begin{align}
  \trace \tup{ \Pi_\delta^n(\sigma_{XB}|x^n) \cN(x^n) }
    &\geq 1-2^{-b'\delta n} ,
  \label{eq:UnitTCond} \\
  2^{-n(H(B|X)_\sigma+c'\delta)} \Pi_\delta^n(\sigma_{XB}|x^n)
   &\preceq  \Pi_\delta^n(\sigma_{XB}|x^n) \,\cN(x^n)\, \Pi_\delta^n(\sigma_{XB}|x^n)
   \nonumber\\ 
   &\preceq 2^{-n(H(B|X)_{\sigma}-c'\delta)} \Pi_\delta^n(\sigma_{XB}|x^n),
  \label{eq:rhonProjIneqCond}
  \\
  \trace( \Pi_\delta^n(\sigma_{XB}|x^n))&\leq 2^{n(H(B|X)_\sigma+c'\delta)}
  \label{eq:PidimCond}
\end{align}
where $b',c'>0$ is a constant, $\cN(x^n) = \bigotimes_{i=1}^n \cN_{X_i \to B_i}(x_i)$.
Furthermore,
\begin{align}
\trace( \Pi_\delta^n(\sigma_B) \, \cN(x^n) )&\geq 1-2^{-c'\delta n} 
\label{eq:UnitTCondB}
\end{align}
(see \cite[Property 15.2.7]{Wilde:17b}),
and hence, by the Gentle Operator Lemma~\cite[Lemma 9.4.2]{Wilde:17b},
\begin{gather}
  \label{eq:gentleProjector}
  \norm[\Big]{ \cN(x^n) 
    - \Pi_\delta^n(\sigma_B) \, \cN(x^n) \, \Pi_\delta^n(\sigma_B)
  }_1 \leq 2^{-c'\delta n/2 + 1}.
\end{gather}

\section{Error Analysis for Single-User Achievability}
\label{sec:achiev.single.error_analysis}

We show that for some \(\tau>0\), the error
probabilities of the random code \(\cC_{\CodeBL}\) satisfy
\begin{subequations}
  \label{eq:capSAID-CBC.direct.single-user}
\begin{align}
  \lim_{n\to\infty}
  \Pr\set[\bigg]{
    \max_{i \in [N]}
      \erra(\cL_{A\to B}, n, \codet_{\CodeBL}, i)
    \geq e^{-n\tau}
  } &= 0,
  \label{eq:capSAID-CBC.direct.single-user.typeA} \\
  \lim_{n\to\infty}
  \Pr\set[\bigg]{
    \max_{i,i' \in [N]} \max_{i \neq i'}
      \errb(\cL_{A\to B}, n, \codet_{\CodeBL}, i', i)
    \geq e^{-n\tau}
  } &= 0.
  \label{eq:capSAID-CBC.direct.single-user.typeB}
\end{align}
\end{subequations}

The codebook that is used here is the same as in the classical derivation
\cite{bracherLapidoth2017idbc,bracher2016PhD}. Hence, we can
use the cardinality bounds for the index bins \(\set{\bcV_i}_{i\in [N]}\)
that were established in
\cite{bracherLapidoth2017idbc,bracher2016PhD}.
Denote the collection of index bins by
\(\bcV^N= \set{\bcV_i}_{i\in [N]}\).
\begin{lemma}[see {\cite[Lemma 5]{bracherLapidoth2017idbc}}]
Given \(\mu > 0\), let \(\cG_\mu\) be the set of all realizations \(\cV^N\) of
\(\bcV^N\) such that
\begin{align}
  \abs{\cV_i} &> (1-\delta_n)e^{n\tR}, \label{eq:Vi_lb} \\
  \abs{\cV_i} &< (1+\delta_n)e^{n\tR}, \label{eq:Vi_ub} \\
  \abs{\cV_i \cap \cV_{i'}} &< 2 \delta_n e^{n\tR} \label{eq:Vii_ub}
\end{align}
for all \(\cV_i, \cV_{i'} \in \cV^N, i \neq i'\), where
\(\delta_n = e^{-n\mu/2}\). Then, the probability that \(\bcV^N \in \cG_\mu\)
converges to 1 as \(n\to\infty\), \ie
\begin{equation}\label{eq:VinG}
  \lim_{n\to\infty} \Pr\set{\bcV^N \in \cG_\mu} = 1,
\end{equation}
for \(\mu < \Rpool - \Rbin\).
\label{lemma:BL-binning.cardinalities}
\end{lemma}
Hence, it suffices to consider the bin collection realizations \(\cV^N\) of
\(\bcV^N\) that satisfy \eqref{eq:Vi_ub}--\eqref{eq:Vii_ub},
for \(\mu \in (0, \Rpool - \Rbin)\).
Thus, the input state is
as in
\eqref{eq:capSAID.proof.enc.single-user}, since
 \(\cV_i \neq \emptyset\) by \eqref{eq:Vi_lb}.

\paragraph{Missed ID Error}
Consider an index bin \(\cV_i \in \cV^N\).
We bound the probability of the missed-ID error (first kind), given by
\begin{align}
  \erra(\cL_{A\to B}, n, \codet_\CodeBL, i)
    &= \frac{1}{\abs{\cV_i}} \sum_{v \in \cV_i}
    \Pr\tup[\Big]{ a(v')=0 ,\,\text{for all $v' \in \cV_{i}$} \,\Big|\, \text{$F(v)$ was sent}}
    \label{eq:capSAID-CBC.missed-ID.expression}
\end{align}
Note that for $\rho_{B^n}^{F(v)} = \cL_{A \to B}(F(v))$,
\begin{align}
  \Pr
    &\tup{ a(v')=0 ,\,\text{for all $v' \in \cV_{i}$} \,\middle|\, \text{$F(v)$ was sent} }
      \nonumber\\
    &= \trace\tup{
        \prod_{v' \in \cV_i} \tup{ \identity - \Pi \Pi^{F(v')} \Pi }
        \rho_{B^n}^{F(v)}
      }
      \nonumber\\
    &\leq \trace\tup{
        \tup[\big]{ \identity - \Pi \Pi^{F(v)} \Pi }
        \, \rho_{B^n}^{F(v)}
      }
      \nonumber\\
                        &= 1 - \trace\tup{
        \Pi^{F(v)} \,
        \Pi \,\rho_{B^n}^{F(v)} \, \Pi
      }
      \nonumber\\
    &\leq 1
       - \trace\tup{ \Pi^{F(v)} \rho_{B^n}^{F(v)} }
       + \norm{ \rho_{B^n}^{F(v)} - \Pi \, \rho_{B^n}^{F(v)} \, \Pi }_1.
\end{align}
By the Gentle Operator Lemma~\cite[Lemma 9.4.2]{Wilde:17b},
we have
$\norm{ \rho_{B^n}^{F(v)} - \Pi \, \rho_{B^n}^{F(v)} \, \Pi }_1 \leq e^{- n c'\delta /2 + 1}$
(see~\eqref{eq:gentleProjector}), because $F(v) \in \cT_\delta^n(P_X)$, for all $v \in \cV$.
Since also $\trace\tup{\Pi^{F(v)} \rho_{B^n}^{F(v)}} \geq 1 - e^{-nb'\delta}$~\eqref{eq:UnitTCond},
there exists $\tau_1 > 0$ such that
\begin{gather}
  \erra(\cL_{A\to B}, n, \codet_\CodeBL, i) \leq e^{-nb'\delta} + e^{- n c'\delta /2 + 1} < e^{-n\tau_1}.
  \label{eq:ranCode.error.idOne.proof}
\end{gather}

\paragraph{False ID Error}

Next, we bound the probability of an error of the second kind.
Suppose that the sender sends an ID message \(i\) and the receiver is
interested in \(i' \neq i\).
Recall that we can restrict our attention to
realizations \(\cV^N= \set{\cV_i} \in \cG_\mu\), following
Lemma~\ref{lemma:BL-binning.cardinalities}.
Let $v\in\cV_i$ be the index that Alice has chosen.
Observe that
\begin{align}
   & \errb (\cL_{A\to B}, n, \codet_\CodeBL, i', i)
      \nonumber\\
     &= \frac{1}{\abs{\cV_i}} \sum_{v \in \cV_i}
        \Pr\tup{ \exists\, v'\in\cV_{i'}: a(v')=1 ~\middle|~ \text{$F(v)$ was sent} }
      \nonumber\\
     &\leq \frac{1}{\abs{\cV_i}}\sum_{v \in \cV_i \cap \cV_{i'}}
        1
             + \frac{1}{\abs{\cV_i}}\sum_{v \in \cV_i \cap \cV_{i'}^c}
        \Pr\tup{ \exists\, v'\in\cV_{i'}: a(v')=1 ~\middle|~ \text{$F(v)$ was sent} }
        \nonumber\\
    &\leq \frac{\abs{\cV_i \cap \cV_{i'}}}{\abs{\cV_i}}
      \nonumber\\ & \quad
       + \frac{1}{\abs{\cV_i \cap \cV_{i'}^c}}\sum_{v \in \cV_i \cap \cV_{i'}^c}
        \Pr\tup{ \exists\, v'\in\cV_{i'}: a(v')=1 ~\middle|~ \text{$F(v)$ was sent} },
    \label{eq:BL-code.errB.split}
\end{align}
since any probability is at most $1$ and
\(\abs{\cV_i \cap \cV_{i'}^c}\leq \abs{\cV_i }\).
The first term is bounded by
\begin{equation}
  \frac{\abs{\cV_i \cap \cV_{i'}}}{\abs{\cV_i}}
  < \frac{2\delta_n}{1-\delta_n}
  < \delta_n
  \label{eq:BL-binning.Vii_ub.pretty}
\end{equation}
(see \eqref{eq:Vi_lb} and \eqref{eq:Vii_ub}), where the second inequality holds as
\(\delta_n < 1/2\), for sufficiently large \(n\).

It remains to bound the second term in the right-hand side of
\eqref{eq:BL-code.errB.split}, for which
\(v\in\cV_i\) and \(v\notin\cV_{i'}\).
For every pair of indices \(v \notin \cV_{i'}\) and
\(v' \in \cV_{i'}\), we have that the codewords
\(F(v)\) and \(F(v')\) are statistically independent.

Assume without loss of generality that 
$\cV_{i'} = \set{ 1,2,\dots,\abs{\cV_{i'}} }$.
Let $\cA_{i'}(v')$ denote the event that $v' \in \cV_{i'}$ is the first index to hit  'yes' as the measurement outcome, \ie
$a(1)=a(2)=\dots=a(v'-1)=0$ and $a(v')=1$.
Then,
\begin{align}
  \Pr \tup{ \cA_{i'}(v') \,\middle|\, \text{$F(v)$ was sent} }
  &= \trace\tup{
      D_{v'} D_{v'-1}^c \cdots D_1^c
      \; \rho^{F(v)}_{B^n}
      D_1^c \cdots D_{v'-1}^c
    }
    \nonumber\\
  &\leq \trace\tup{ D_{v'} \; \rho^{F(v)}_{B^n} }
    \nonumber\\
      &= \trace\tup{ \Pi^{F(v')} \Pi \; \rho^{F(v)}_{B^n} \; \Pi }
\end{align}
where $D_v = \Pi \Pi^{F(v)} \Pi$, $D_v^c = \identity - D_v$
and again $\rho_{B^n}^{F(v)} = \cL_{A \to B}(F(v))$.
Thus, we have \begin{align}
  \expect_\CodeBL
  &\intv{
    \Pr\tup{ \cA_{i'}(v') \,\middle|\, \text{$F(v)$ was sent} }
    }
    \leq \trace\tup{ \expect \intv{\Pi^{F(v')}} \Pi \; \expect \intv{\rho^{F(v)}_{B^n}} \Pi },
\end{align}
and note that by \eqref{eq:probTypicalLB}, there exists $\epsilon_n$ with $\lim_{n \to \infty} = 0$ such that
\begin{align}
  \expect \intv{\rho^{F(v)}_{B^n}}
  &
  = \frac{1}{P_X(\cT_\delta^n(P_X))} \sum_{x^n \in \cT_\delta^n(P_X)} P_X^n(x^n) \rho^{x^n}_{B^n}
        = \frac{1}{1 - \epsilon_n} \rho_B^{\otimes n}.
\end{align}
It follows that
\begin{IEEEeqnarray}{rcl}
  (1-\epsilon_n) \expect_\CodeBL
  \intv[\big]{
    \Pr\tup{ \cA_{i'}(v') \,\middle|\, \text{$F(v)$ was sent} }
    }
    &  = & 
  \trace\tup{ \expect \intv{\Pi^{F(v')}} \Pi \; \rho_B^{\otimes n} \; \Pi }
  \nonumber\\ &
    \overset{(a)}{\leq} &
      e^{-n(H(B)_\rho - c\delta)}
      \trace\tup{ \expect \intv{\Pi^{F(v')}} \Pi }
        \nonumber\\ &
    \overset{(b)}{<} &
      e^{-n(H(B)_\rho - H(B|X)_\rho - (c+c')\delta)}
  \nonumber\\ &
    = & e^{-n(I(X;B)_\rho - (c+c')\delta)},
\end{IEEEeqnarray}
where (a) holds by \eqref{eq:rhonProjIneq}, and (b) by \eqref{eq:PidimCond}
since $F(v') \in \cT_\delta^n(P_X)$ for all $v' \in \cV$.
Therefore, there exists \(\theta > 0\) such that
\begin{IEEEeqnarray}{rcl}
  &\expect_\CodeBL
  &
  \intv[\big]{
      \Pr\tup{ \exists\, v' \in \cV_{i'} : a(v') \,\middle|\, \text{$F(v)$ was sent} }
    }
  \nonumber\\
  &=&
  \expect_\CodeBL
  \intv[\big]{
      \Pr\tup{ \exists\, v' \in \cV_{i'} : \cA_{i'}(v') \,\middle|\, \text{$F(v)$ was sent} }
    }
  \nonumber\\ &
    \overset{(a)}{\leq}  &
    \frac{1}{1-\epsilon_n}
    \sum_{v' \in \cV_{i'}} \expect_\CodeBL \intv{
      \Pr\tup{ \cA_{i'}(v') \,\middle|\, \text{$F(v)$ was sent} }
    }
  \nonumber\\ &
    \leq &
    \frac{1}{1-\epsilon_n}
    \abs{\cV_{i'}} e^{-n(I(X; B) - (c+c')\delta)}
  \nonumber\\ &
    \overset{(b)}{<} &
    \frac{1+\delta_n}{1-\epsilon_n}
    e^{-n(I(X; B)_\rho - \Rbin - (c+c')\delta)}
  \nonumber\\ &
    \overset{(c)}{<} &
    e^{-n\theta},
  \label{eq:BL-code.errB.expect.indep}
\end{IEEEeqnarray}
where
(a) follows from the union-of-events bound,
(b) is due to \eqref{eq:Vi_ub},
and
(c) holds since $(c+c')\delta < I(X;B)_\rho - \Rbin$ by \eqref{eq:deltaBound}.

We show now that the false-ID error is small with high probability.
Let \(\alpha\) satisfy
\begin{equation}
  0 < \alpha < (\Rbin - R)/2.
  \label{eq:alphaRbin}
\end{equation}
By the union-of-events bound,
\begin{align}
  \Pr &\tup{
    \max_{i' \in [N]} \max_{i \neq i'}
      \errb(\cL_{A\to B}, n, \codet_{\CodeBL}, i', i)
      \geq \delta_n + e^{-n\theta} + e^{-n\alpha}
    }
    \nonumber\\
  &\leq \sum_{i' \in [N]} \sum_{i \neq i'}
    \Pr\tup{
      \errb(\cL_{A\to B}, n, \codet_{\CodeBL}, i', i)
      \geq \delta_n + e^{-n\theta} + e^{-n\alpha}
    }.
    \label{eq:BL-code.errB.boundA}
\end{align}
Note that by \eqref{eq:BL-code.errB.split} and \eqref{eq:BL-binning.Vii_ub.pretty},
\begin{multline}
  \errb (\cL_{A\to B}, n, \codet_\CodeBL, i', i)
    \\
  \leq \delta_n
    + \frac{1}{\abs{\cV_i \cap \cV_{i'}^c}} \sum_{v \in \cV_i \cap \cV_{i'}^c}
    \Pr\tup{ \exists\, v'\in\cV_{i'}: a(v')=1 \,\middle|\, \text{$F(v)$ was sent} }.
\end{multline}
Therefore, there exists \(\tau_2 > 0\) such that
\begin{align}
  \Pr
  &\set{
      \max_{i' \in [N]} \max_{i \neq i'}
      \errb(\cL_{A\to B}, n, \codet_{\CodeBL}, i', i)
      \geq e^{-n\tau_2}
    }
    \nonumber\\
  &\leq \sum_{i' \in [N]} \sum_{i \neq i'}
    \Pr\set[\bigg]{
      \frac{1}{\abs{\cV_i \cap \cV_{i'}^c}} \sum_{v \in \cV_i \cap \cV_{i'}^c}
      \Pr\tup{ \exists\, v'\in\cV_{i'}: a(v')=1 \,\middle|\, \text{$F(v)$ was sent} }
    \nonumber\\ &\hspace{8em}
      \geq e^{-n\theta} + e^{-n\alpha}
    }
    \nonumber\\
  &\overset{(a)}{\leq}  \exp\tup{-2 e^{-2n\alpha} \abs{\cV_i \cap \cV_{i'}^c}} \nonumber\\
  &\overset{(b)}{<}  \exp\tup{2e^{nR} - e^{-2n\alpha} e^{-n\Rbin}}
      , \label{eq:ranCode.error.idTwo.proof}
\end{align}
for sufficiently large \(n\), where (a) follows from
Hoeffding's inequality~\cite[Theorem~1]{hoeffding1963inequalities},
since the codewords \(F(v), v \in \cV\) are i.i.d., (b) follows from
\(N= \exp \tup{e^{nR}}\), and
\begin{gather}
  \abs{\cV_i \cap \cV_{i'}^c}
  = \abs{\cV_i} - \abs{\cV_i \cap \cV_{i'}}
  > (1-\delta_n) e^{n\Rbin} - 2\delta_n e^{n\Rbin}
  \geq e^{n\Rbin}/2,
\end{gather}
as \(\abs{\cV_i} \geq (1-\delta_n) e^{n\Rbin}\) and
\(\abs{\cV_i \cap \cV_{i'}} < 2\delta_n e^{n\Rbin}\), by Lemma \ref{lemma:BL-binning.cardinalities}
(see \eqref{eq:Vi_lb} and \eqref{eq:Vii_ub}, respectively),
where the last inequality follows from \(\delta_n < 1/2\),
for sufficiently large \(n\).

Based on 
\eqref{eq:ranCode.error.idOne.proof}
and \eqref{eq:ranCode.error.idTwo.proof}, we
have established that \eqref{eq:capSAID-CBC.direct.single-user}
holds for \(\tau = \min\set{ \tau_1,\tau_2}\).

\section{Proof of Inequality (\ref{eq:ConvBR1})}

The rate bound is based on the combinatorial argument in the single-user converse proof, due to Ahlswede and Winter \cite{AhlswedeWinter2002quantumID}.
First, we  define a quantum hypergraph and give the Ahlswede-Winter covering lemma \cite{AhlswedeWinter2002quantumID}.
\begin{definition}
A quantum hypergraph $(\cH,\cG)$ is defined by a finite-dimentional Hilbert space $\cH$ and a finite collection  $\cG=\{G_x \}_{x\in\cX}$ of operators on $\cH$, where $0\preceq G_x\preceq \identity$ for $x\in\cX$.
\end{definition}

\begin{lemma}[{see \cite[Lemma 9]{AhlswedeWinter2002quantumID}}]
\label{lemm:qHypergraph}
Let $(\cH,\cG)$ be a quantum hypergraph such that $G\preceq \eta\identity$ for all $G\in\cG$, and fix $\epsilon,\tau>0$. Given a probability distribution $P$ on $\cG$, define
\begin{align}
    \rho=\sum_{G\in\cG} P(G) G \,.
\end{align}
Then, there exists a subspace $\cH_0\subseteq \cH$ and operators $G_1,\ldots,G_L\in\cG$ such that
\begin{align}
    &\trace(\Pi_0\rho)\leq \tau\\
    &(1-\epsilon)\Pi_1\rho\Pi_1 \preceq \Pi_1 \bar{\rho} \Pi_1\preceq (1+\epsilon)\Pi_1\rho\Pi_1\\
    & L\leq 1+\eta|\cH|\frac{2\ln 2 \log(2|\cH|)}{\epsilon^2 \tau}
\end{align}
with
\begin{align}
    \overline{\rho}=\frac{1}{L}\sum_{\ell=1}^L G_\ell,
\end{align}
while $\Pi_0$ and $\Pi_1$ are the orthogonal projections onto $\cH_0$ and $\cH_1\equiv \cH/\cH_0$, respectively.
\end{lemma}
Intuitively, the covering lemma has the following  interpretation. We can think of $\rho$ as the average of an ensemble $\cG$ of operators.
Then, we consider a compressed ensemble, $\{G_1,\ldots,G_L\}\subseteq \cG$, from which an operator is drawn uniformly. The resulting average of this compression is $\overline{\rho}$. Then, within the subspace $\cH_1$,
the projection 
$\Pi_1 \rho \Pi_1$ onto $\cH_1$ is almost unaffected by the compression.
The idea in  the single-user converse proof for the  classical-quantum channel $\channel_{X\to B_1}^{(1)}$, is now to replace the arbitrary distributions $Q_{i_1}$  of an
ID code by uniform distributions $\overline{Q}_{i_1}$
on subsets of $\cA_{P^*}$, with cardinality bounded by $L\approx e^{n\mathsf{I}(P^*;\channel^{(1)})}$. The condition is that the corresponding output
states are close, so the resulting ID code will have similar error probabilities. 
As reliable identification requires the encoder to assign a different distribution $\overline{Q}_{i_1}$ to each message $i_1$, the number messages is thus bounded by the number of options for choosing $L$ input sequences, which is  $|\cX^n|^L$.
Now, we formalize this argument. Let $\lambda>0$ be arbitrarily small, such that  $\lambda<1-\lambda_1^*-\lambda_2^*$.
Let $\Pi^{\delta}(B_1^n)$ and $\Pi_\delta^n(B_1^n|x^n)$ denote the $\delta$-typical projector and the conditional $\delta$-typical projector, respectively, for
$\rho_{B_1}=\sum_{x\in\cX} P^*(x)\channel_{A\to B_1}^{(1)}(x)$.
Then, denote the projection of each output state onto the $\delta$-typical subspace by
\begin{gather}
    \Upsilon_{X^n\to B_1^n}(x^n)=
    \Pi^{2\delta}(B_1^n) \; \Pi_\delta^n(B_1^n|x^n)  \;  \channel^{(1)}_{X^n\to B_1^n}(x^n)
     \;  \Pi_\delta^n(B_1^n|x^n)  \; \Pi^{2\delta}(B_1^n).
\end{gather}
Then, for sufficiently large $n$,
\begin{align}
    \left\lVert   \Upsilon(x^n)-\channel^{(1)}(x^n)  \right\rVert_1\leq \frac{\lambda}{6} ,
    \label{eq:UpsilonDistN}
\end{align}
for all $x^n\in\cA_{P^*}$,
by \cite[Lemma V.9]{Winter:99p}.

We apply Lemma~\ref{lemm:qHypergraph} with $\epsilon=\tau=\frac{\lambda^2}{1200}$ to the quantum hypergraph for which the vertex space $\cH$ is the range of $\Pi^{2\delta}(B_1^n)$, and  the edges are $\Upsilon_{X^n\to B_1^n}(x^n)$, $x^n\in \cA_{P^*}$. 
Thereby, for every $i_1\in [N_1']$, there exist $L_1$ sequences $x^n(\ell_{i_1})\in \cA_{P^*}$,
\begin{align}
    L_1=e^{n(\mathsf{I}(P^*;\channel^{(1)}_{A\to B_1})+\delta_n)},
\end{align}
  such that 
$G_{\ell_{i_1}}=\Upsilon_{X^n\to B_1^n}(x^n(\ell_{i_1}))$ satisfy the properties in the lemma, with
$\overline{\rho}_{B_1^n}^{i_1}=\frac{1}{L_1}\sum_{j_{i_1}=1}^{L_1} \Upsilon_{X^n\to B_1^n}(x^n(j_{i_1}))$. Then, the uniform $L_1$-distribution $\overline{Q}_{i_1}^{P^*}$ satisfies
\begin{align}
    \left\lVert Q_{i_1}^{P^*} \Upsilon^{(k)}-\overline{Q}_{i_1}^{P^*} \Upsilon^{(k)} \right\rVert_1 \leq \frac{\lambda}{6}.
\end{align}
Thus, by (\ref{eq:UpsilonDistN}),
\begin{align}
    \left\lVert Q_{i_1}^{P^*} \channel^{(1)}-\overline{Q}_{i_1}^{P^*} \channel^{(1)} \right\rVert_1 \leq \frac{\lambda}{3} 
\end{align}
for $i_1\in [N_1']$.
Therefore, $\{\overline{Q}_{i_1}^{P^*},D_{i_1}^{(1)}\}$ is an $(n,N_1',\lambda_1^*+\frac{\lambda}{3},\lambda_2^*+\frac{\lambda}{3})$ code for the single-user channel $\channel^{(1)}_{A\to B_1}$.

Since $\lambda_1^*+\lambda_2^*+\frac{2\lambda}{3}<1$, each message must have a different input distribution. That is, $\overline{Q}_{i_1}^{P^*}\neq \overline{Q}_{i_1'}^{P^*}$ for $i_1\neq i_1'$. As each $\overline{Q}_{i_1}^{P^*}$ is uniform over $L_1$ sequences in $\cX^n$, it follows that
\begin{align}
    N_1'
    &\leq |\cX^n|^{L_1}
    \nonumber\\
    &=\exp( e^{n(\mathsf{I}(P^*;\channel^{(1)}_{A\to B_1})+\delta_n+\frac{1}{n}\log \log|\cX|)})
    \nonumber\\
    &\leq \exp(e^{n(\mathsf{I}(P^*;\channel^{(1)}_{A\to B_1})+\epsilon_1)})
\end{align}
for sufficiently large $n$. Inequality (\ref{eq:ConvBR1}) readily follows.
\pushQED

\section*{Declarations}
The authors have no relevant financial or non-financial interests to disclose.

A part of this paper was presented at the
2022 IEEE International Symposium on Information Theory
\cite{PeregRosenbergerDeppe2022id_quantum_BC_isit}.
\ifarxiv\else
  A preprint of this work was posted on arXiv
  \cite{PeregRosenbergerDeppe2022id_quantum_BC_arxiv}.
\fi

Data sharing not applicable to this article as no datasets were generated or
analysed during the current study.

\printbibliography

\balance

\end{document}